\documentclass[twocolumn, twocolappendix]{aastex631}

\usepackage[utf8]{inputenc}
\usepackage{graphicx}
\graphicspath{{./}{./CurrentFigures/}{./tex/figures/}}
\usepackage{amsmath}
\usepackage{afterpage}
\usepackage{enumitem}
\usepackage{xcolor}
\usepackage{float}
\usepackage{tabularx}
\usepackage{multirow}
\setenumerate{itemsep=0mm}%noitemsep}

\usepackage{float}

\definecolor{newblue}{cmyk}{1,0.7,0,0}
\definecolor{newa}{cmyk}{2,0.3,1,0}

\usepackage{soul} 
\usepackage{amsmath}
\usepackage{amssymb}
\usepackage{xspace}
\usepackage{xifthen}
\usepackage{eso-pic}

\newcommand{\smbhmass}{M_{\rm{BH}}}

\newcommand{\stellarmass}{M_{\star}}

\newcommand{\dmbhmstar}{$\Delta\log\left( \smbhmass / \stellarmass \right)$\xspace}

\newcommand{\Mbh}{$\smbhmass$\xspace}

\newcommand{\Mstar}{$\stellarmass$\xspace}

\newcommand{\massrelation}{$\smbhmass$\textendash \Mstar}

\newcommand{\quasarhost}{{quasar\textendash host}\xspace}
\newcommand{\agnhost}{{AGN\textendash host}\xspace}

\newcommand{\mgii}{{MgII}}
\newcommand{\civ}{{CIV}}
\newcommand{\hbeta}{{H$\beta$}}

\newcommand{\one}    {SDSS\,J1004$+$4112\xspace}
\newcommand{\ones}    {SJ1004\xspace}
\newcommand{\two}    {SDSS\,J1029$+$2623\xspace}
\newcommand{\twos}    {SJ1029\xspace}
\newcommand{\three}     {SDSS\,J2222$+$2745\xspace}
\newcommand{\threes}     {SJ2222\xspace}
\newcommand{\four}     {SDSS\,J0909$+$4449\xspace}

\newcommand{\five}     {SDSS\,J1326$+$4806\xspace}
\newcommand{\fives}     {SJ1326\xspace}
\newcommand{\six}     {COOL\,J0542$-$2125\xspace}

\newcommand{\seven}     {COOL\,J0335$-$1927\xspace}
\newcommand{\sevens}     {CJ0335\xspace}
\newcommand{\eight}     {COOL\,J1153$+$0755\xspace}
\newcommand{\eights}     {CJ1153\xspace}

\newcommand{\hst}{\textit{HST}\xspace}
\newcommand{\jwst}{\textit{JWST}\xspace}

\newcommand{\numpy}{\code{NumPy}}

\newcommand{\matplotlib}{\code{matplotlib}}
\newcommand{\astropy}{\code{Astropy}}

\newcommand{\galfit}{{GALFIT}\xspace}
\newcommand{\prospector}{\code{Prospector}}
\newcommand{\dynesty}{\code{dynesty}}
\newcommand{\lenstool}{\code{Lenstool}}
\newcommand{\dsnine}{{SAO Image DS9}\xspace}
\newcommand{\pyspeckit}{\code{PySpecKit}}

\definecolor{forestgreen}{HTML}{228B22}
\definecolor{urlblue}{HTML}{000000}

\newcommand{\ie}{i.e.\xspace}
\newcommand{\eg}{e.g.\xspace}

\newcommand{\etal}{et al.\xspace}

\mathchardef\mhyphen="2D

\newcommand{\less}{\ensuremath{ {<}\,} }
\newlength{\dhatheight}

\newcommand{\code}[1]{\texttt{#1}\xspace}

\newcommand{\unit}[1]{\ensuremath{\mathrm{\,#1}}\xspace}
\newcommand{\yr}{\unit{yr}}

\newcommand{\angstrom}{\unit{\AA}}

\newcommand{\km}{\unit{km}}
\newcommand{\kms}{\km \second^{-1}}

\newcommand{\kpc}{\unit{kpc}}
\newcommand{\second}{\unit{s}}

\newcommand{\Msun}{M_\odot}
\newcommand{\Msunyr}{\Msun\yr^{-1}}

\newcommand{\e}{\unit{e^{-}}}

\newcommand{\secref}[1]{Section~\ref{sec:#1}}
\newcommand{\appref}[1]{Appendix~\ref{app:#1}}
\newcommand{\tabref}[1]{Table~\ref{tab:#1}}

\newcommand{\figref}[1]{Figure~\ref{fig:#1}}

\newcommand{\eqnref}[1]{Equation~\eqref{eqn:#1}}

\newcommand{\bandvar}[2][]{%
  \ifthenelse{\isempty{#1}}{\var{#2}}{\var{#2\_#1}}%
}

\newcommand{\SExtractor}{\code{SExtractor}}
\newcommand{\sextractor}{\SExtractor}

\newcommand{\emcee}{\code{emcee}}

\newcommand{\var}[1]{\ensuremath{\texttt{\MakeUppercase{#1}}}\xspace}

\providecommand\physrep{\ref@jnl{Phys.~Rep.}}%
\providecommand\apjs{\ref@jnl{ApJS}}%
\providecommand{\jcap}{\ref@jnl{JCAP}}%
\usepackage[T1]{fontenc}

\newcommand{\wslqmodel}{{0.41\pm 0.31}}
\newcommand{\wslqmodelsel}{{-0.42\pm 0.31}}
\newcommand{\litagnmodel}{{2.04\pm 0.19}}
\newcommand{\litagnmodelsel}{{1.04\pm 0.21}}
\newcommand{\selwslq}{{0.43}}

\begin{document}

\title{COOL-LAMPS VIII: Known wide-separation lensed quasars and their host galaxies reveal a lack of evolution in $M_{\rm{BH}}/M_\star$ since $z\sim 3$}

\author[0000-0001-9978-2601]{Aidan P. Cloonan}
\affiliation{Department of Astronomy and Astrophysics, University of Chicago, 5640 South Ellis Avenue, Chicago, IL 60637, USA}
\affiliation{Kavli Institute for Cosmological Physics, University of Chicago, 5640 South Ellis Avenue, Chicago, IL 60637, USA}
\affiliation{Department of Astronomy, University of Massachusetts, 710 North Pleasant Street, Amherst, MA 01003, USA}

%\author[0000-0002-3475-7648]{Gourav Khullar (\Punj{ਗੌਰਵ ਖੁੱਲਰ})}
\author[0000-0002-3475-7648]{Gourav Khullar}
\affiliation{Department of Physics and Astronomy and PITT PACC, University of Pittsburgh, Pittsburgh, PA 15260, USA}

\author[0000-0003-4470-1696]{Kate A. Napier}
\affiliation{Department of Astronomy, University of Michigan, 1085 S. University Ave, Ann Arbor, MI 48109, USA}

\author[0000-0003-1370-5010]{Michael D. Gladders}
\affiliation{Department of Astronomy and Astrophysics, University of Chicago, 5640 South Ellis Avenue, Chicago, IL 60637, USA}
\affiliation{Kavli Institute for Cosmological Physics, University of Chicago, 5640 South Ellis Avenue, Chicago, IL 60637, USA}

\author[0000-0003-2200-5606]{H{\aa}kon Dahle}
\affiliation{Institute of Theoretical Astrophysics, University of Oslo, P.O. Box 1029, Blindern, NO-0315 Oslo, Norway}

\author[0000-0001-7905-2134]{Riley Rosener}
\affiliation{Department of Astronomy and Astrophysics, University of Chicago, 5640 South Ellis Avenue, Chicago, IL 60637, USA}

\author[0009-0000-9780-4328]{Jamar Sullivan Jr.}
\affiliation{Department of Astronomy and Astrophysics, University of Chicago, 5640 South Ellis Avenue, Chicago, IL 60637, USA}

\author[0000-0003-1074-4807]{Matthew B. Bayliss}
\affiliation{Department of Physics, University of Cincinnati, Cincinnati, OH 45221, USA}

\author[0009-0005-1143-495X]{Nathalie Chicoine}
\affiliation{Department of Astronomy and Astrophysics, University of Chicago, 5640 South Ellis Avenue, Chicago, IL 60637, USA}

\author[0000-0003-0896-8502]{Isaiah Escapa}
\affiliation{Department of Astronomy and Astrophysics, University of Chicago, 5640 South Ellis Avenue, Chicago, IL 60637, USA}

\author[0009-0003-0226-6988]{Diego Garza}
\affiliation{Department of Astronomy and Astrophysics, University of Chicago, 5640 South Ellis Avenue, Chicago, IL 60637, USA}
\affiliation{Department of Astronomy and Astrophysics, University of California, Santa Cruz, CA 95064, USA}

\author[0009-0007-1440-1832]{Josh Garza}
\affiliation{Department of Astronomy and Astrophysics, University of Chicago, 5640 South Ellis Avenue, Chicago, IL 60637, USA}

\author[0000-0001-9816-0878]{Rowen Glusman}
\affiliation{Department of Astronomy and Astrophysics, University of Chicago, 5640 South Ellis Avenue, Chicago, IL 60637, USA}
\affiliation{Institute of Physics, University of Amsterdam, Science Park 904, 1098 XH Amsterdam, Netherlands}

\author[0000-0003-2294-4187]{Katya Gozman}
\affiliation{Department of Astronomy, University of Michigan, 1085 S. University Ave, Ann Arbor, MI, 48109-1107, USA}

\author[0009-0006-6950-6351]{Gabriela Horwath}
\affiliation{Department of Astronomy and Astrophysics, University of Chicago, 5640 South Ellis Avenue, Chicago, IL 60637, USA}

\author[0009-0006-7664-877X]{Andi Kisare}
\affiliation{Department of Astronomy and Astrophysics, University of Chicago, 5640 South Ellis Avenue, Chicago, IL 60637, USA}

\author[0000-0001-8000-1959]{Benjamin C. Levine}
\affiliation{Department of Astronomy and Astrophysics, University of Chicago, 5640 South Ellis Avenue, Chicago, IL 60637, USA}
\affiliation{Department of Physics and Astronomy, Stony Brook University, 100 Nicolls Rd, Stony Brook, NY 11794, USA}

\author{Olina Liang}
\affiliation{Department of Astronomy and Astrophysics, University of Chicago, 5640 South Ellis Avenue, Chicago, IL 60637, USA}

\author[0000-0002-5825-7795]{Natalie Malagon}
\affiliation{Department of Astronomy and Astrophysics, University of Chicago, 5640 South Ellis Avenue, Chicago, IL 60637, USA}

\author[0000-0002-8397-8412]{Michael N. Martinez}
\affiliation{Department of Physics, University of Wisconsin, Madison, 1150 University Avenue, Madison, WI 53706, USA}

\author[0000-0002-3361-2893]{Alexandra Masegian}
\affiliation{Department of Astronomy and Astrophysics, University of Chicago, 5640 South Ellis Avenue, Chicago, IL 60637, USA}
\affiliation{Department of Astronomy, Columbia University, 538 West 120th Street, New York, NY 10027, USA}

\author[0000-0001-9225-972X]{Owen S. Matthews Acu\~{n}a}
\affiliation{Department of Astronomy, University of Wisconsin, Madison, Madison, WI 53706, USA}

\author[0000-0002-5573-9131]{Simon D. Mork}
\affiliation{Department of Astronomy and Astrophysics, University of Chicago, 5640 South Ellis Avenue, Chicago, IL 60637, USA}

\author[0009-0001-5944-5624]{Kunwanhui Niu}
\affiliation{Department of Astronomy and Astrophysics, University of Chicago, 5640 South Ellis Avenue, Chicago, IL 60637, USA}

\author[0000-0002-2862-307X]{M. Riley Owens}
\affiliation{Department of Physics, University of Cincinnati, Cincinnati, OH 45221, USA}

\author[0000-0002-7922-9726]{Yue Pan}
\affiliation{Department of Astronomy and Astrophysics, University of Chicago, 5640 South Ellis Avenue, Chicago, IL 60637, USA}
\affiliation{Department of Astrophysical Sciences, Princeton University, 4 Ivy Ln, Princeton, NJ 08544, USA}

\author[0000-0002-7627-6551]{Jane R. Rigby}
\affiliation{Astrophysics Science Division, Code 660, NASA Goddard Space Flight Center, 8800 Greenbelt Rd., Greenbelt, MD 20771, USA}

\author[0000-0002-7559-0864]{Keren Sharon}
\affiliation{Department of Astronomy, University of Michigan, 1085 S. University Ave, Ann Arbor, MI 48109, USA}

\author[0000-0002-2323-303X]{Isaac Sierra}
\affiliation{Department of Astronomy and Astrophysics, University of Chicago, 5640 South Ellis Avenue, Chicago, IL 60637, USA}

\author[0000-0002-2718-9996]{Antony A. Stark}
\affiliation{Center for Astrophysics | Harvard \& Smithsonian, 60 Garden St, Cambridge, MA 02138, USA}

\author[0000-0002-1106-4881]{Ezra Sukay}
\affiliation{Department of Physics and Astronomy, Johns Hopkins University, 3400 N. Charles Street, Baltimore, MD 21218, USA}

\author[0009-0008-0518-8045]{Marcos Tamargo-Arizmendi}
\affiliation{Department of Astronomy and Astrophysics, University of Chicago, 5640 South Ellis Avenue, Chicago, IL 60637, USA}

\author[0000-0001-6584-6144]{Kiyan Tavangar}
\affiliation{Department of Astronomy, Columbia University, 538 West 120th Street, New York, NY 10027, USA}

\author[0000-0002-5279-0230]{Raul Teixeira}
\affiliation{Department of Astronomy and Astrophysics, University of Chicago, 5640 South Ellis Avenue, Chicago, IL 60637, USA}

\author[0009-0008-6557-2065]{Kabelo Tsiane}
\affiliation{Department of Astronomy and Astrophysics, University of Chicago, 5640 South Ellis Avenue, Chicago, IL 60637, USA}

\author[0000-0003-0295-875X]{Grace Wagner}
\affiliation{Department of Astronomy and Astrophysics, University of Chicago, 5640 South Ellis Avenue, Chicago, IL 60637, USA}

\author[0000-0002-6779-4277]{Erik A. Zaborowski}
\affiliation{Department of Astronomy and Astrophysics, University of Chicago, 5640 South Ellis Avenue, Chicago, IL 60637, USA}
\affiliation{Department of Physics, The Ohio State University, Columbus, OH 43210, USA}
\affiliation{Center for Cosmology and Astro-Particle Physics, The Ohio State University, Columbus, OH 43210, USA}

\author[0000-0001-6454-1699]{Yunchong Zhang}
\affiliation{Department of Astronomy and Astrophysics, University of Chicago, 5640 South Ellis Avenue, Chicago, IL 60637, USA}
\affiliation{Department of Physics and Astronomy and PITT PACC, University of Pittsburgh, Pittsburgh, PA 15260, USA}

\author[0009-0006-4143-1159]{Yifan ``Megan'' Zhao}
\affiliation{Department of Astronomy and Astrophysics, University of Chicago, 5640 South Ellis Avenue, Chicago, IL 60637, USA}

\submitjournal{\apj}

\correspondingauthor{Aidan Cloonan}
\email{apcloonan@umass.edu}

\shorttitle{Wide-separation lensed quasars and their host galaxies}
\shortauthors{A. P. Cloonan \etal}

\begin{abstract}

Wide-separation lensed quasars (WSLQs) are a rare class of strongly lensed quasars, magnified by foreground massive galaxy clusters, with typically large magnifications of the multiple quasar images. They are a relatively unexplored opportunity for detailed study of quasar host galaxies. The current small sample of known WSLQs has a median redshift of $z\approx 2.1$, larger than most other samples of quasar host galaxies studied to date. Here, we derive precise constraints on the properties of six WSLQs and their host galaxies, using parametric surface brightness fitting, measurements of quasar emission lines, and stellar population synthesis of host galaxies in six WSLQ systems.  Our results, with significant uncertainty, indicate that these six hosts are a mixture of star-forming and quiescent galaxies. To probe for co-evolution between AGNs and host galaxies, we model the offset from the `local' ($z=0$) $M_{\rm{BH}}$\textendash$M_\star$ relation as a simple power-law in redshift. Accounting for selection effects, a WSLQ-based model for evolution in the $M_{\rm{BH}}$\textendash$M_\star$ relation has a power-law index of $\gamma_M=-0.42\pm0.31$, consistent with no evolution. Compared to several literature samples, which mostly probe unlensed quasars at $z<2$, the WSLQ sample shows less evolution from the local relation, at $\sim4\sigma$. We find that selection effects and choices of $M_{\rm{BH}}$ calibration are the most important systematics in these comparisons. Given that we resolve host galaxy flux confidently even from the ground in some instances, our work demonstrates that WSLQs and highly magnified AGNs are exceptional systems for future AGN\textendash{host} co-evolution studies.

\end{abstract}


%-------------------------------------------------------------------------------

\section{Introduction}
\label{sec:intro}

At their gravitational centers, galaxies contain supermassive black holes (SMBHs), the mass (\Mbh) of which is known to correlate with other physical properties of galaxies, such as the stellar mass \citep[\Mstar; e.g.,][]{Magorrian_1998, Haring_2004}, the stellar velocity dispersion of the bulge \citep[e.g.,][]{Ferrarese_2000}, and the host galaxy luminosity \citep[e.g.,][]{Marconi_2003}. These relations reflect an underlying cosmological history of co-evolution between SMBHs and their host galaxies \citep[e.g.,][]{Kormendy_2013}. Accretion of matter from the surrounding galaxy onto the SMBH fuels outflows of gas and radiation,
which we observe from various inclination angles as active galactic nuclei (AGNs), according to the `unified model' \citep{Antonucci_1993, Netzer_2015, Ogawa_2021}. From a theoretical perspective, due to the highly energetic nature of accretion-powered outflows, we might anticipate a co-evolutionary relationship between AGNs and galaxies. Indeed, theoretical works have continuously hypothesized an `AGN feedback' scenario, in which AGN activity plays a significant role in regulating star formation activity in massive galaxies \citep{Bower_2006, Croton_2006, Hopkins_2006, Somerville_2008, Habouzit_2021, Wellons_2023}.

From an observational perspective, the details of the co-evolutionary relationship between AGNs and their host galaxies are poorly understood \citep[for some reviews, see][]{Fabian_2012, King_2015, Padovani_2017, Veilleux_2020, Sajina_2022}. Unobscured quasars\textemdash viewed at an angle such that the accretion disk is directly visible\textemdash are extremely luminous \citep{LyndenBell_1969} and can outshine their host galaxies by orders of magnitude in the optical and near-infrared (NIR) regime \citep{Hopkins_2006}. This large relative difference in surface brightness significantly limits the ability to spatially resolve and study their host galaxies even at low redshift \citep{Bahcall_1997}. We effectively observe distant quasars as point sources.

Disentangling quasar flux from host flux requires high signal-to-noise (S/N) and/or resolution. Since unobscured quasars are very luminous in the UV through mid-IR, millimeter observations such as from ALMA provide better contrast of molecular outflows and cold gas in the host galaxy against radiative flux from the quasar \citep[e.g.,][]{Omont_1996, Wagg_2012, Venemans_2017}. Combining integral field spectroscopy and adaptive optics has significantly improved ground-based mapping of quasar outflows \citep{Kakkad_2023}. Furthermore, despite persistent observational difficulties in PSF modeling and \quasarhost decomposition \citep{Zhuang_2023}, space-based observatories have allowed for much clearer detection and study of quasar host galaxies. \hst observations of quasars at lower redshift ($z \lesssim 0.5$) have found that the host galaxies vary in morphology from early- to late-type galaxies \citep{Dunlop_2003, Guyon_2006} and that they can have regions with large star formation rates \citep{Young_2014}. \hst additionally facilitated the first cosmological studies of co-evolution between AGNs and host galaxies over redshift \citep{Peng_2006, Merloni_2010}. Recently, early studies with \jwst are probing gas, star formation, and accretion-driven outflows around distant quasars \citep{Ding_2022a, Kocevski_2023, Cresci_2023} and further surpassing previous observational constraints to study AGNs and their environments near the end of cosmic reionization \citep{Furtak_2023, Marshall_2023, Larson_2023, Pacucci_2023, Wang_2023} and even earlier to $z\sim 10$ \citep{Bogdan_2023, Goulding_2023}.

The aforementioned scaling relations between SMBH and host galaxy properties were established using relatively local samples \citep[$z \lesssim 0.1$;][]{Haring_2004}. Extending these scaling relations to high redshift to study cosmic evolution is an ongoing problem \citep[e.g.,][]{Suh_2020, Ding_2020, Li_2023, Pacucci_2023, Tanaka_2024}. Not only are host galaxies even more difficult to observe, but the masses of quasar SMBHs are often constrained with secondary scaling relations derived from quasar emission \citep[e.g.,][]{Wandel_1999, Vestergaard_2002, Mejia_2016} rather than a physically motivated model. Reverberation mapping reduces uncertainty in SMBH mass measurements significantly, but is almost entirely restricted to unobscured AGNs at $z < 1$ \citep{Li_2023}. Selection bias is another enormous challenge. Since AGN luminosity is explicitly related to SMBH mass \citep{Marconi_2004}, all flux- or luminosity-limited sample selections inevitably bias the parameter space(s) of interest \citep{Lauer_2007, Schulze_2011}.

Strong gravitational lensing of background quasars provides several opportunities to overcome the above challenges. Notably, it provides effective spatial resolution of quasar host galaxies at $z > 1$ in optical and IR imaging \citep{Bayliss_2017}. While a lensed quasar still appears as a point source, the lens stretches and magnifies the host galaxy light, separating quasar and host galaxy flux on the sky. 

Nearly all known strongly lensed quasars\textemdash several hundred in total\textemdash are lensed by single galaxies \citep{Lemon_2023}. Angular image separations between the lens galaxy and lensed quasar in these systems are typically $\sim 1\arcsec$. There are eight known lensed quasars with image separations greater than 10\arcsec, each lensed by a massive galaxy cluster \citep{Inada_2003, Inada_2006, Dahle_2013, Shu_2018, Shu_2019, Martinez_2023, Napier_2023a, Kisare_2024}. We refer to these systems as \textit{wide-separation lensed quasars} (WSLQs).
Previous studies have used quasars that are strongly lensed by galaxies to study \quasarhost co-evolution over cosmic time, as far back as redshift $z \sim 4.5$ \citep[e.g.,][]{Peng_2006, Stacey_2020, Ding_2021}.
However, strong lensing by galaxy clusters provides much higher magnification and larger image separation, better separating the flux from the quasar and host galaxy than galaxy-scale lensing does. Recent studies of WSLQs, motivated by the larger magnification and better natural resolution, have focused on radio emission \citep{McKean_2021, Hartley_2021}, quasar variability and structure \citep{Williams_2021a, Williams_2021b, Hutsemekers_2023, Fian_2024}, and cosmology \citep{Napier_2023b}.

In this work, using archival and new imaging and spectroscopy from \hst, Keck Observatory, the Magellan Telescopes, and the Nordic Optical Telescope (NOT), we study the quasar and host galaxy properties in six of the known WSLQ systems, for which we have reliable detections of host galaxy flux. (The remaining two are not detected in extant data.) This WSLQ sample spans a redshift range of $1.5 < z < 3.3$. With constraints on the physical properties, we model the \massrelation relation and study its evolution with redshift, as a probe of co-evolution between AGNs and galaxies, while accounting for selection bias and other systematics.

This paper is organized as follows. In \secref{data}, we outline a combination of archival data and new observations used for analysis of each WSLQ. In \secref{analysis}, we describe methodologies for modeling quasar and galaxy properties.
\secref{results} discusses the \massrelation relation and the implications of our modeling results. In particular, we directly compare to samples at different cosmological epochs from the literature. Then, \secref{systematics} investigates several important sources of bias to qualify the key measurements. We discuss takeaways and physical interpretations of these various analyses in \secref{discussion}. Finally, we summarize our work and our conclusions in \secref{summary}.

Magnitudes are reported in the AB system. For all geometric and cosmological calculations, we assume a flat $\Lambda$CDM cosmology with $\Omega_{\Lambda}$ = 0.7, $\Omega_{m}$ = 0.3, and $H_{0} = 70\ \rm{km}\,\rm{s}^{-1}\,\rm{Mpc}^{-1}$.

\section{Targets and Observational Data}
\label{sec:data}

\begin{figure*}
    \hspace*{-1.5cm}
    \includegraphics[width=21cm]{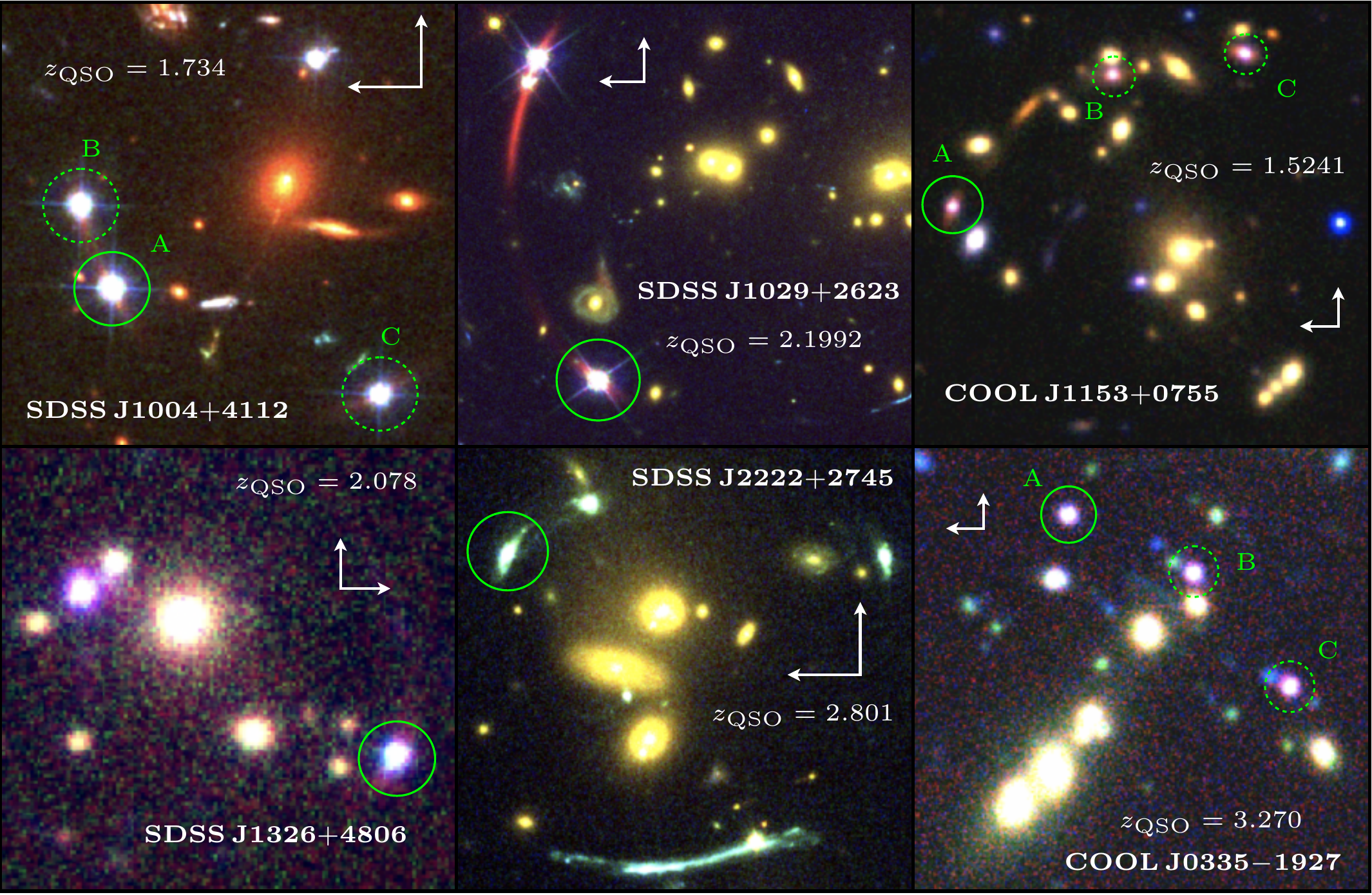}
    \caption{RGB images of each WSLQ system. The quasar images analyzed in this work are circled in green, with primary images labeled as `A' or with a solid circle. For the WSLQs where we later measure the properties of multiple images, the additional quasar images are shown with dotted circles. Each quasar redshift and line-of-sight name is given in the figure. A scale of 3\arcsec\ is given in each cutout with a pair of white arrows marking the North and East directions (where North is always clockwise of East).}
    \label{fig:wslq_images}
\end{figure*}

The name of each WSLQ is given in \tabref{data}, and images are shown in \figref{wslq_images}. For brevity, in the rest of this paper, we abbreviate each WSLQ with an initial and four numbers, \eg \one as \ones.

We use a combination of archival and new observations in this work. To carry out any robust analysis of a galaxy hosting a luminous AGN, a detection (with convincingly high signal-to-noise) of extended flux beyond the point source of an AGN is required. While \hst imaging easily detects flux from WSLQ host galaxies, only three known WSLQs have previously existing \hst imaging \citep{Inada_2005, Oguri_2013, Sharon_2017}, with a fourth in preparation (GO-17243, PI: Napier) and a fifth to be visited in the coming months (GO-17431, PI: Gladders).
Host galaxy detection with ground-based observations at optical wavelengths requires longer exposures \citep{Martinez_2023}, a challenge which guides observation planning and selection of archival data. Each of the six WSLQs and a summary of all of the observational data used for analysis is listed in \tabref{data}.

\begin{table*}
%\centering
\hspace*{-2.5cm}
%\begin{tabular}{p{8em} p{4em} p{4em} p{2.5em} p{2.5em} p{18em} p{5.5em} p{9em}}
\begin{tabular}{c c c c c c c c}
 \hline
 \multirow{2}{4em}{\centering{Name}} & \multirow{2}{3.5em}{\centering{RA}} & \multirow{2}{3.5em}{\centering{Dec}} 
 & \multirow{2}{2.5em}{\centering $z_{\rm{lens}}$} & \multirow{2}{2.5em}{\centering $z_{\rm{QSO}}$} 
 & \multirow{2}{15em}{\centering Imaging Data} & \multirow{2}{6.0em}{\centering Spectroscopic Data} & \multirow{2}{9em}{\centering Discovery Reference}\\ [3.5ex]
 \hline
 \ones & 151.1429 & 41.2118 & 0.68 & 1.734 & \multirow{2}{15em}{\centering \hst F435W, F555W, F814W, F160W; \textit{Spitzer} IRAC Ch1\textendash 4} & eBOSS & \cite{Inada_2003} \\ [3ex] 
 \twos & 157.3077 & 26.3917 & 0.584 &  2.1992\footnote{Updated redshift from \cite{Acebron_2022a}.} & \hst F475W, F814W, F160W & Keck LRIS & \cite{Inada_2006} \\
 \threes & 335.5354 & 27.7594 & 0.49 & 2.801\footnote{Updated redshift from \cite{Acebron_2022b}.} & \hst F435W, F606W, F814W, F160W & n/a\footnote{\threes has a measurement for \Mbh from reverberation mapping \citep{Williams_2021a, Williams_2021b}} & \cite{Dahle_2013} \\
 \fives & 201.5001 & 48.1121 & 0.396 & 2.078 & NOT $gHK_S$  & eBOSS & \cite{Shu_2019} \\
 \sevens & 53.7701 & $-$19.4688 & 0.4178 & 3.27 & Magellan $grzH$ & Magellan LDSS3-C & \cite{Napier_2023a} \\
 \eights & 178.3302 & 7.9325 & 0.42 & 1.5241 & Magellan $grzJHK_S$ & Magellan LDSS3-C & \cite{Kisare_2024} \\
\end{tabular}
\caption{\label{tab:data} The sample of wide-separation lensed quasars and the data used for each. Redshifts are taken from the given discovery reference for \ones and \fives, and from \cite{Acebron_2022a, Acebron_2022b} for \twos and \threes respectively. Imaging of \fives is from the Nordic Optical Telescope (NOT), while imaging and spectroscopy of \sevens and \eights are from the Magellan Telescopes.}
\end{table*}

\subsection{Archival Observations}
\label{archival_data}

For photometry of each of \ones, \twos, and \threes, we use \hst WFC3 and/or ACS imaging in the observed-frame optical and NIR. \hst imaging came through GO-10509 (PI: Kochanek), GO-9744 (PI: Kochanek), and GO-10793 (PI: Gal-Yam) for \ones; through GO-12195 (PI: Oguri) for \twos; and through GO-13337 (PI: Sharon) for \threes. The SDSS extended Baryon Oscillation Spectroscopic Survey (eBOSS) provides calibrated spectra of resolution $R\sim 2000$ \citep{Smee_2013, Dawson_2016} for \ones, \twos, and \fives. \twos was also observed on December 15, 2007 (Program C245L, PI: Ofek) with the Low Resolution Imaging Spectrometer (LRIS) at Keck Observatory \citep{Oke_1995}. We use this archival spectrum for \twos as it has higher signal-to-noise (S/N) than eBOSS, and we use eBOSS spectra for \ones and \fives. We do not perform spectroscopic analysis of \threes, as there exists a robust measurement of \Mbh via reverberation mapping \citep{Williams_2021a, Williams_2021b}.

We carry out wavelength- and flux-calibrations for LRIS spectroscopy of \twos using standard IRAF tools \citep{Tody_1986}. The LRIS spectrum is flux-calibrated to the corresponding eBOSS spectrum of the object to ensure accuracy and minimize uncertainty in the calibration process overall. This results in a calibrated spectrum of eBOSS's resolution but with higher S/N.

\subsection{New Observations}
\label{sec:new_data}

We conducted observations of \fives with NOT and of \sevens and \eights with the Magellan Telescopes. Below, we describe observing for each; in particular, we also briefly introduce the \eights system, as it contains a previously unreported WSLQ. We performed all reductions, including image stacking and spectral flux/wavelength calibrations, using IRAF \citep{Tody_1986}.

\subsubsection{\five}

We have a calibrated optical spectrum of \fives from eBOSS. From NOT imaging programs for host galaxy photometry, we detect host galaxy flux in $g$, $H$, and $K_S$ filters. The $g$-band detection is part of an ongoing monitoring program with NOT/ALFOSC to measure time-delays between quasar images for cosmological study. The total exposure time is 14290s, from a stack of 56 single-epoch images. Then, we observed \fives in $K_S$ on April 27, 2021 with NOT/NOTCAM for 3600s of total exposure as part of Program 63-017 (PI: Dahle). Observations in $H$-band were taken on March 3, 2023 as part of Program 66-011 (PI: Dahle), with an exposure time of 3564s.

\subsubsection{\seven}

As in \cite{Napier_2023b}, we use photometric and spectroscopic observations from Magellan Clay/LDSS3-C, taken on September 18, 2022 and on Feb 1, 2023. The total combined integration times from both nights were 1080s in the $g$ and $r$ filters and 720s in the $z$ filter. We use the same reduced $grz$ imaging in this paper. To measure flux around the Balmer break (rest-frame 4000 \AA), we observed of \sevens in $H$-band with Baade/FOURSTAR on December 10, 2022, with a total integration time of 1310s.

\subsubsection{\eight}

\eights is a previously unreported WSLQ system, the most recently discovered by the COOL-LAMPS collaboration \citep{Khullar_2021} as a continuation of its lensed quasar search \citep{Martinez_2023, Napier_2023b}. We briefly describe the system here, but we will defer to an upcoming paper \cite{Kisare_2024} for a more detailed analysis of its properties. The \eights line of sight includes a massive galaxy cluster\textemdash detected in a joint X-ray\textendash Sunyaev-Zeldovich survey by \cite{Tarrio_2019}\textemdash which strongly lenses a group of galaxies at $z \sim 1.5$. This lensed group includes the quasar-hosting galaxy \eights. The quasar, lensed group, lensing cluster and other various lensed sources will be imaged by \hst in Cycle 31 Program GO-17431 (PI: Gladders).

Imaging in $grzJHK_S$ filters and spectroscopy of \eights were collected over the course of several Magellan nights in early 2023: February 1, February 6, March 9, and March 24. The total exposure time for $grz$ imaging (Clay/LDSS3-C; Feb. 1 and Mar. 24) was 540s per filter, while for $JHK_S$ (Baade/FOURSTAR; Feb. 6 and Mar. 9) we obtained total exposure times of 3429s for $J$ and 3493s each for $H$ and $K_S$. Optical spectroscopy for each quasar image was acquired via Magellan Clay/LDSS3-C (Feb. 1 and Mar. 24).

\section{Measurements and Analysis}
\label{sec:analysis}

\subsection{Quasar\textendash Host Photometry}
\label{sec:photometry}

\begin{figure}
    %\hspace*{-0.45cm}
    \centering
    \includegraphics[width=\columnwidth]{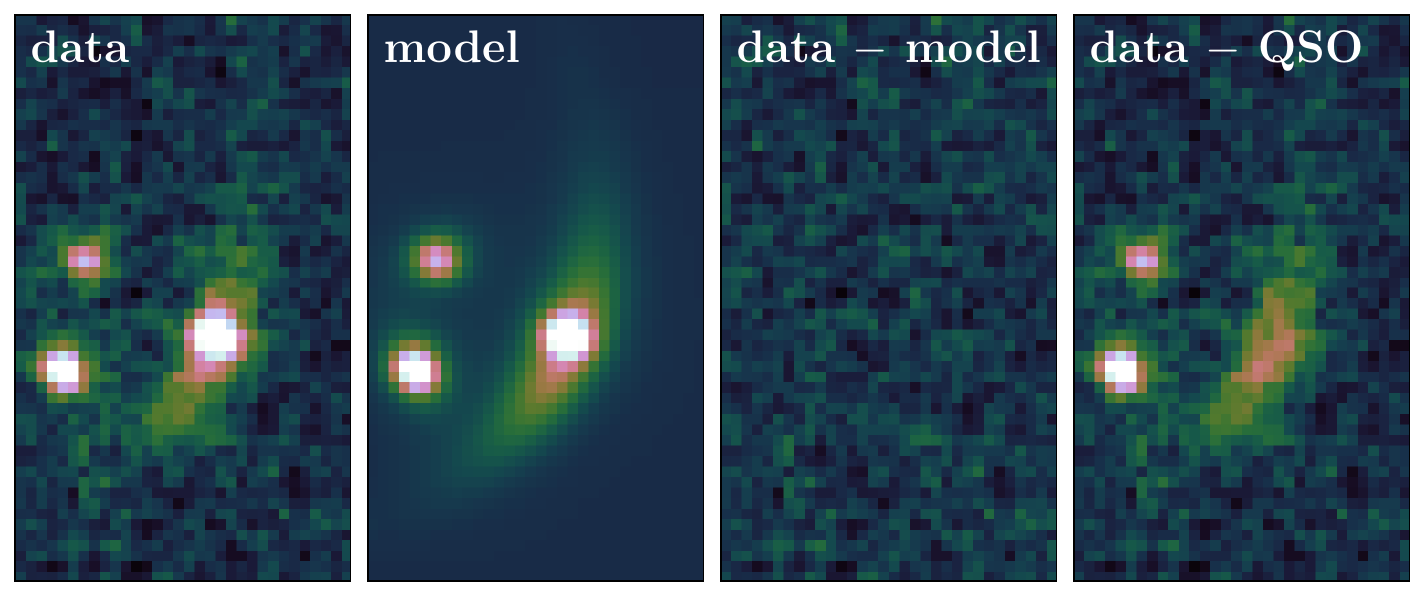}
    \caption{\galfit SB model of \fives in the $K_S$ filter, observed with NOT. From left to right: reduced $K_S$ image, SB model fit to the image, model residual, reduced image with model quasar subtracted. Note that this decomposition appears sufficient for measuring host galaxy photometry even with ground-based imaging.}
    \label{fig:galfit}
\end{figure}

To constrain host galaxy properties, we begin by constructing surface brightness (SB) profiles of the lensed quasars and their host galaxies using \galfit \citep{GALFIT_2002, GALFIT_2010}. \galfit performs parametric modeling of SB profiles for astronomical sources. The code allows for least-squares fitting of complex light profiles consisting of multiple sources or components. We carry out \quasarhost decomposition by simultaneously measuring quasar and host magnitudes and fitting nearby contaminants along the line of sight. An example \galfit model is shown in \figref{galfit}. For each WSLQ, the process for photometry with \galfit is as follows:
\begin{enumerate}
    \item We construct a model PSF for each imaging filter. (A more detailed explanation is given in the next paragraph.)
    \item With a PSF as an input, for each filter, we optimize a \galfit model of the line of sight around the quasar image. The quasar and foreground stars are each fit with a single point source, while host galaxy arcs, lensing cluster galaxies, and other contaminating sources are fit with one or more S\'ersic profiles.
    \item We quantify noise in the image by computing a root-mean-square (RMS) map of the optimized model residual. (i.e. image minus \galfit model, such as in the third panel of \figref{galfit}.) This RMS map is then smoothed by a 2D Gaussian kernel, which is symmetric with a size of 3$\sigma$.
    \item We generate an array of mock images by multiplying each of 1000 unitary random Gaussian fields by the smoothed RMS map, and then adding the optimized \galfit model.
    \item Finally, we propagate uncertainty on the host galaxy flux measurement by re-optimizing the \galfit model on each of these 1000 mock fields, providing a probability distribution function for host galaxy flux.
\end{enumerate}

PSF modeling for \hst imaging is different from that for ground-based imaging; we briefly summarize the two techniques here. For an \hst image, we identify a set of stars of `medium' brightness, i.e., those which neither oversaturate the CCDs nor are very faint. The images of stars are stacked, and each input image is azimuthally filtered about the center and at larger radii, to minimize contamination from nearby sources. Then, we optimize a basic S\'ersic model for each diffraction spike (with the stacked image subtracted) and then add the spike models to the stacked image, giving an empirical PSF.
For a ground-based image, we directly construct a model PSF by fitting one or more Moffat profiles \citep{Moffat_1969} to a non-saturated reference star in the line of sight. 

For ground-based images of \fives, \sevens, and \eights, we empirically measure magnitude zeropoints in order to properly calibrate quasar and host galaxy magnitude measurements into the AB system. Either manually or using \sextractor \citep{sextractor_paper}, we carry out aperture photometry on bright stars near the line of sight, convert the resulting flux into a magnitude, and then measure the zeropoint by adding the resulting magnitude to an existing calibrated measurement in a catalog. We use 2MASS catalogs \citep[which have Vega magnitudes, requiring a final filter-dependent translation to AB;][]{Skrutskie_2006} for NIR filters and archival DESI Legacy Surveys for optical filters.

Our method with \galfit provides calibrated flux measurements of the quasars and host galaxies in the image plane. In the strong gravitational lensing regime, the demagnified flux in the source plane is directly proportional to the flux in the image plane as $f_{\rm{source}} = \mu_{\rm{m}} f_{\rm{image}}$, where $\mu_{\rm{m}}$ is the magnification coefficient for the lensed image. To demagnify flux into the source plane, we use magnification coefficients reported in \cite{Napier_2023b, Napier_2023a} for \ones, \twos, \threes, and \sevens. For \fives and \eights, we use magnifications computed from ground-based lens models \citep[to be described by][]{Kisare_2024} following the procedure of \cite{Napier_2023a}, which are less precise than models based on \hst imaging \citep{Sharon_2017}. All lens models and magnification maps were constructed using the lens modeling software \lenstool \citep{lenstool_2007}.

\subsection{Stellar Population Synthesis Modeling}
\label{sec:spsmodeling}

We employ stellar population synthesis (SPS) to constrain host galaxy stellar masses and spectral energy distributions (SEDs).
% Description of a `delayed-tau' model
Using \prospector \citep{prospector_paper}, we implement a parametric treatment of star formation using a `delayed-tau' model, where star formation rate (SFR) follows a `delayed' exponential decay as a function of time, \ie, ${\rm{SFR}}(t) \propto t e^{-t/\tau}$ where $\tau$ is a free parameter.
For these parametric models, \prospector samples in the parameter space of mass, age, and the folding timescale $\tau$.

While SPS provides remarkably powerful tools for inferring physical properties of unresolved stellar populations, we note two critical caveats in the approach. First and foremost, in galaxy SED modeling, one usually has to make critical assumptions about the star formation history (SFH). The choice of parametric SFH may bias stellar mass measurements by up to $0.2 - 0.5$ dex \citep{Carnall_2019}, depending on the model and on the type of galaxy (star-forming/quiescent). The second is the age-metallicity-dust degeneracy \citep[e.g.,][]{Conroy_2013}. 
These problems are well known, and we must look to minimize their impacts on our measurements. We discuss SED modeling systematics in \secref{sps_sys}.

As outlined previously, the number of photometric filters in which the host galaxy is robustly detected varies from system to system. With that in mind, we approach SPS modeling for each WSLQ system differently in an attempt to minimize systematic bias. For all stellar population modeling, we assume a Chabrier initial mass function \citep[IMF;][]{Chabrier_2003} and the dust attenuation law from \cite{Calzetti_2000}.

Depending on how well a model constrains the stellar mass, we change the array of free parameters in a given SPS model, either including additional free parameters or fixing parameters. At the very least, we look to constrain total stellar mass $M_{\star, \rm{tot}}$, galaxy age $t_{\rm age}$, $e$-folding timescale $\tau$, and the optical depth from diffuse interstellar dust $\tau_{\lambda,2}$ with each SED template. While we prefer to change a model slightly to align with what a galaxy's photometry can capably constrain, we recognize that this heterogeneity in modeling can introduce bias in key inferred parameters. We include a simple test for this concern in \secref{sps_sys}. Note that in the formalism described above, the remnant stellar mass corresponds to the galaxy mass contained in stars, once corrected foor mass loss via stellar winds and supernova feedback. The term `stellar mass' $\stellarmass$ refers specifically to the remnant stellar mass throughout the rest of this paper.

Below, we outline our model choices and setup for each quasar host galaxy.
With the exception of \ones, we use \galfit photometry as described in \secref{photometry}.

\textbf{\one}.
\cite{Ross_2009} compute and report demagnified flux values for the host galaxy of \ones in 8 total imaging filters, four from \hst (optical and NIR) and four from \textit{Spitzer} IRAC (mid-IR). This \textit{HST}$+$\textit{Spitzer} imaging remains the highest quality data available for this WSLQ. As such, we use the fluxes from \cite{Ross_2009} and recalculate the demagnified flux in each filter using the updated lens model from \cite{Napier_2023b}.

In addition to $M_{\star, \rm{tot}}$ and $t_{\rm age}$, we fit for $\tau$, $\tau_{\lambda,2}$, and stellar metallicity $\log(Z / Z_\odot)$. Given the larger constraining power from 8 photometric filters, we include NIR dust emission (as a nuisance parameter), nebular continuum, and line emission in the model. The dust emission comes from the heating of dust grains by rest-UV photons. Physically, the nebular emission arises from interstellar gas, which is characterized by an ionization parameter $U$ and gas-phase metallicity $Z_{\rm{gas}}$. We let ionization $U$ be a free parameter and keep gas-phase metallicity fixed at $\log(Z_{\rm{gas}} / Z_{\odot}) = 0$.
Finally, we also introduce a free `constant' parameter $C$ to the model for \ones, which corresponds to a fraction of the total stellar mass formed by a `tophat' term in the ${\rm{SFR}}(t)$ function. In other words, we assume for \ones that SFR follows:
\begin{equation}
    \label{eqn:sfrconst}
    {\rm{SFR}}(t) \propto t \exp \left\{ -\frac{t}{\tau} \right\} + \frac{ C M_{ \star, {\rm{tot}} } }{t_{\rm age}},
\end{equation}
where $t_{\rm age}$ and $M_{\star, {\rm{tot}}}$ are free parameters in the model.

\textbf{\two}.
We have available imaging in three \hst filters: F475W, F814W, and F160W. While this allows for far less flexibility in an SED template for \twos than for \ones, the flux measurement in F160W offers an important constraint because it probes redward of the Balmer break at $z\sim 2$. As with \ones, we cite the image magnifications from the lens model built by \cite{Napier_2023b}.

We use the delayed-tau model, with $M_{ \star, {\rm{tot}}}$, $t_{\rm{age}}$, and $\tau$ as free parameters. We fix the metallicity in stars to $Z_\odot$ and impose an upper limit $\tau_{\lambda,2} < 1$ on the diffuse dust attenuation. The metallicity does not significantly impact the stellar mass measurement in this instance because the model fits the SED amplitude but not elemental abundances. Furthermore, an initial fit with a more lenient prior $\tau_{\lambda,2} < 3$ produces a dusty best-fit with $\log(\stellarmass / \Msun) > 12$ and $\log(\rm{SFR} / \Msunyr) \sim 3$. We suspect this result is likely inaccurate as it implies an overly high SFR for such a massive galaxy \citep[e.g.,][]{Somerville_2008, Leja_2022}, and we therefore assume that the host galaxy is not highly obscured.

\textbf{\three}. 
We use \hst imaging in the F435W, F606W, F814W, and F160W filters and a similar SPS formalism as for \twos. As we have improved constraining power with four filters and we also do not find an unphysical solution, we do not impose the stricter upper limit on $\tau_{\lambda,2}$. Due to its apparent bluer color, we also introduce nebular emission and continuum similarly to \ones. We reuse the magnification computed by \cite{Napier_2023b}.

\textbf{\five}. 
The SED modeling of this galaxy follows an identical process to \twos.
The photometric filters from NOT in which we confidently detect the quasar host galaxy\textemdash $g$, $H$, and $K_S$\textemdash span a similar wavelength range, and the galaxy appears red in those colors as shown in \figref{wslq_images}. A model galaxy with strong dust attenuation is again overly massive and highly star-forming, and so we again assume impose an upper limit of $\tau_{\lambda,2} < 1$, keeping $M_{ \star, {\rm{tot}}}$, $\tau_{\lambda,2}$, $t_{\rm{age}}$, and $\tau$ as free parameters.
The image magnifications are determined from a simpler lens model based on the available ground-based imaging, using a similar methodology as in \cite{Martinez_2023} and \cite{Napier_2023a}.

\textbf{\seven}.
We have one more photometric detection ($grzH$) than for \twos and \fives, and we do not find any possibly unphysical best-fits with a lenient prior on dust attenuation.
As such, we do not impose the upper limit on $\tau_{\lambda,2}$. The metallicity is kept constant at the solar value. Similarly to \ones, we include nebular continuum and emission as well as dust emission. We use ground-based magnification estimates of each quasar image from \cite{Napier_2023b}.

\textbf{\eight}. We model this quasar host galaxy using the same procedure as for \sevens, except with metallicity $\log(Z / Z_\odot)$ as an additional free parameter. Our set of photometric filters is similar but larger ($grzJHK_S$), and so it better samples the galaxy SED. We compute magnifications from a ground-based lens model, which will be discussed in further detail by \cite{Kisare_2024}.

\begin{figure}
    \hspace*{-1.25cm}
    \centering
    \includegraphics[width=285px]{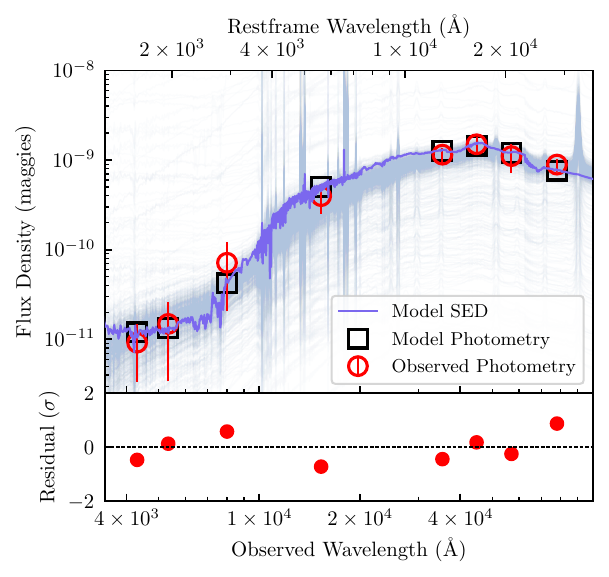}
    \caption{SED model for the host galaxy of \ones. The light-blue band is a set of random samples from the output sampling chain give by the SPS framework. On the bottom panel we show the residuals for each filter, where we define the residual $\sigma = (m_{\rm{obs}} - m) / s_m$, \ie, observed magnitude minus model magnitude divided by observed magnitude uncertainty.}
    \label{fig:sps_1004host}
\end{figure}

\subsection{Analysis of Quasar Spectroscopy}
\label{sec:quasarspec}

For distant AGNs, current methods for constraining \Mbh require three key components: a measurement of velocity dispersion from the unresolved `broad-line region' (BLR), a constraint on the BLR's physical size, and an assumption that the classical virial theorem applies \citep{Peterson_1993}. 
If we measure the BLR radius $R_{\rm{BLR}}$ and its Doppler shift $\Delta V$, then we find a SMBH mass \citep{Williams_2021a, Williams_2021b}:
\begin{equation}
    \label{eqn:smbhvirial}
    \smbhmass = f \frac{R_{\rm{BLR}} \Delta V^2}{G},
\end{equation}
where $f$ is a fudge constant accounting for unknown geometry and kinematics in the BLR, and $G$ is the universal gravitational constant. In lieu of reverberation mapping, empirical correlations between $R_{\rm{BLR}}$ and both emission and continuum luminosities have allowed for \Mbh constraints from a single spectrum \citep{Wandel_1999, Vestergaard_2002}. Although subject to additional uncertainty, these `single-epoch’ measurements require far less observational resources and have allowed for efficient \Mbh measurements for large samples of quasars in surveys \citep[e.g.,][]{Shen_2011, Koss_2022}.

We measure the quasar SMBH mass \Mbh via single-epoch spectroscopy of each WSLQ system except for \threes, for which we cite an existing \Mbh value from reverberation mapping \citep{Williams_2021a, Williams_2021b}.

\subsubsection{Quasar Broad Emission Lines}
\label{sec:specmeasure}

For each quasar spectrum, using \pyspeckit, we fit both the emission profile from the \mgii\ broadline at $\lambda_{\rm rest} = 2799.5\,\angstrom$ and the surrounding continuum with a Voigt profile and a fifth-order polynomial, respectively. This fit provides a measurement of the \mgii\ full-width-at-half-maximum (FWHM), a measurement of velocity dispersion and a continuum luminosity $\lambda L_{\lambda}$, which correlates with $R_{\rm{BLR}}$. 
\pyspeckit is a package designed to replicate spectroscopic tools from IRAF in Python. Its core functions use a basic likelihood maximization scheme, which does not provide uncertainties on free parameters as a Markov Chain Monte Carlo (MCMC) would. However, uncertainties in the measurements of \Mbh are dominated by intrinsic scatter in the luminosity\textendash BLR relations we use, and so we treat the uncertainty contribution from the emission line fitting as negligible.

Regarding the choice of emission line, common choices are \civ\xspace and \mgii\xspace in the rest-UV and the Balmer lines H$\alpha$ and \hbeta\xspace in the rest-optical. We prefer \mgii\xspace to \civ\xspace because the latter's shape is more intrinsically impacted by outflowing, blueshifted gas, leading to systematic bias in measurements of the virialized quasar structure \citep{Mejia_2016}. Generally speaking, the broad Balmer lines are preferred to both \mgii\xspace and \civ\xspace \citep{Shen_2016}, but they are not available with (observed-frame) optical spectroscopy at redshifts $z >1.5$.

\subsubsection{Measuring SMBH Masses}
\label{sec:measure_mbh}

Given the measurements of intrinsic quasar luminosity and of rotational velocity structure described above, we adopt the framework outlined by \cite{Trakhtenbrot_2012} to compute constraints on SMBH masses \Mbh. Assuming that the virial theorem applies (see \eqnref{smbhvirial}), we compute \Mbh via their relation between the continuum luminosity $L_{3000}\equiv 3000\,\angstrom \cdot L_{\lambda}$, the FWHM of \mgii, and \Mbh:
\begin{align}
\begin{aligned}
	\label{eqn:mbh_L3000}
	\smbhmass (L_{3000}) = 5.60 \times 10^{6} \left[ \frac{ L_{3000} }{10^{44}\,\rm{erg\,s}^{-1}} \right]^{0.62} \\
	\times \left[ \frac{ \Delta V_{\rm{\mgii}} }{10^{3}\,\kms} \right]^{2} \Msun, \\
\end{aligned} \\
	%\label{eqn:velocity_width}
	{\rm{where}} \hspace{1cm} \Delta V_{\rm{\mgii}} = \frac{ {\rm{FWHM}} \times c }{ \lambda_{\rm{rest}}(1+z) } \; \kms.
\end{align}

\begin{figure}
    \hspace*{-0.75cm}
    \centering
    \includegraphics[width=275px]{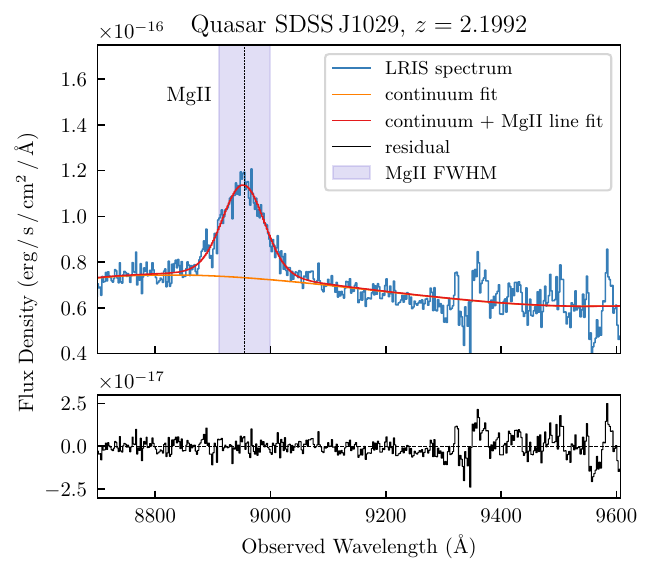}
    \caption{The spectrum surrounding the \mgii\xspace broad emission line of the lensed quasar \twos. Note that the residual plot here is merely (data $-$ model).}
    \label{fig:qsospectrum_1029}
\end{figure}

\cite{Trakhtenbrot_2012} found the above relation by calibrating \mgii\xspace measurements to those via \hbeta\xspace for SDSS quasars. Regarding uncertainties, we simply adopt the scatter associated with Equation \ref{eqn:mbh_L3000} of $0.33\;\rm{dex}$, as they empirically computed.
As a visualization for this process, we show the spectroscopic analysis for \twos in \figref{qsospectrum_1029}.

To measure \Mbh for \sevens, where we do not have access to \mgii, we instead use the \civ\xspace emission line. However, we compute a correction to the FWHM by fitting two Gaussian components to the emission and measuring the centroid blueshift, using the method outlined by \cite{Coatman_2017}. This correction depends on this blueshift, which is a measurement of the centroid's offset from the expected \civ\xspace wavelength at $z = 3.270$. This correction may not bypass the aforementioned bias with \civ-based measurements induced by blueshifted ionized gas, but it still notably reduces the intrinsic scatter in the calibration \citep{Coatman_2017}.

\section{Results}
\label{sec:results}

Having measured the kinematics of quasar BLRs and constructed composite stellar populations for each host galaxy, we have constrained quasar SMBH masses and host galaxy stellar masses. With these in hand, we now investigate the \massrelation relation and compare the WSLQ host galaxies to other samples of AGN-hosting galaxies in the literature. We then combine the WSLQ host galaxies with these other samples and model the change in $\smbhmass / \stellarmass$ as a function of redshift. As is always true for galaxy evolution studies, we have no access to a longitudinal sample, and instead compare time-ordered cross-sectional samples to deduce evolutionary effects \citep[e.g.,][]{Abramson_2016}. Additionally, we should note that SMBHs and galaxies only grow in mass over time, and hence shift to larger values over time in the \massrelation plane. Consequentially, AGN samples at different redshifts but over similar mass ranges are not necessarily antecedents/descendants of one another. 

\subsection{\massrelation Relation}
\label{sec:massrelation}

The \massrelation relation illuminates the relative growth of SMBHs and their host galaxies. For a point of comparison, we adopt the nearby AGNs at $z < 0.1$ from \cite{Bennert_2015, Bennert_2021} as a representative sample of the local relation. For self-consistency, we recalibrate the SMBH masses to the \cite{Trakhtenbrot_2012} formalism using the \hbeta\xspace emission line properties and \href{https://academic.oup.com/mnras/article/427/4/3081/972532\#16748362}{Eq. 3} from \cite{Trakhtenbrot_2012}, and we convert their stellar masses to a Chabrier IMF from a Kroupa IMF by subtracting $0.075\,\rm{dex}$, a conversion that \cite{Bennert_2021} use in their own analysis. We supplement this sample with the local non-active SMBH sample from \cite{Haring_2004}. Following the literature \citep[e.g.,][]{Peng_2006, Kormendy_2013, Li_2023}, we adopt a power law relation between SMBH masses and host galaxy stellar masses for the local sample at $z \sim 0$, following:
\begin{equation}
	\label{eqn:mass_powerlaw}
	\log{\left( {\frac{\smbhmass} {\Msun} } \right)} = \alpha_{1} + \beta_{1} \log{\left( {\frac{\stellarmass} {10^{10} \Msun} } \right)} + n(\sigma_{\rm{int}}),
\end{equation}
where $\alpha_1$ is the normalization (intercept), $\beta_1$ is the slope, and $\sigma_{\rm{int}}$ is the intrinsic scatter in the local relation. The function $n(\sigma_{\rm{int}})$ represents random Gaussian deviates centered at $0$ with scale $\sigma_{\rm{int}}$. 

\begin{figure}
    \hspace*{-0.6cm}
    \centering
    \includegraphics[width=265px]{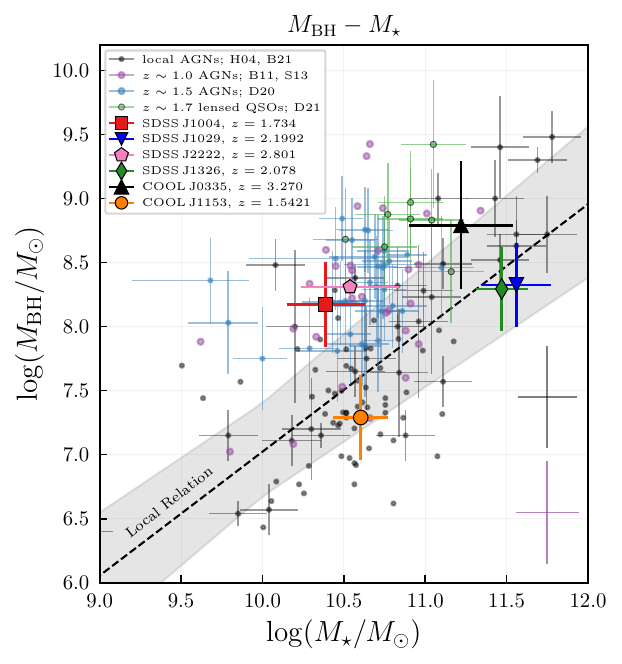}
    \caption{The WSLQ hosts plotted in \massrelation parameter space against samples at other redshifts from the literature. The grey datapoints are the sample of low-redshift AGN hosts at $z \less 0.1$, collated from \cite{Haring_2004} and \cite{Bennert_2021}. The purple datapoints are intermediate-redshift AGN hosts at $0.5 < z < 2$, with a median redshift of $z \sim 1$, from \cite{Bennert_2011b} and \cite{Schramm_2013}. The blue datapoints are AGN hosts from D20, at $1.2 < z < 1.7$ with a median $z \sim 1.5$, and the magenta datapoints are several lensed quasars from \cite{Ding_2021}. The black and purple error bars in the lower right-hand corner correspond to the uncertainties on the \cite{Bennert_2021} sample and the intermediate-$z$ AGNs, respectively. The dotted black line is the best log-linear fit to the local sample at $z < 0.1$, with the shaded grey region corresponding to the $\pm 1\sigma$ fit from the Monte Carlo chain.}
    \label{fig:mass_plot_basic}
\end{figure}

To provide a $\pm 1\sigma$ constraint on the \massrelation at $z \sim 0$, we model the above line via an MCMC scheme and find $\alpha_1 = 7.02\pm 0.38$, $\beta_1 = 0.97\pm 0.11$, and $\sigma_{\rm{int}} = 0.37\pm 0.05$ for the combined local sample. \figref{mass_plot_basic} plots the local relation data and model against the WSLQs and several AGN samples at redshift $z > 0.5$\textemdash a sample of $1 < z < 2$ AGNs from \cite{Bennert_2011b}; one of \textit{Chandra} X-ray-selected $0.5 < z < 1.1$ AGNs \citep{Schramm_2013}; non-lensed AGNs at $1.2 < z < 1.7$ studied by \cite{Ding_2020}, henceforth denoted as D20; and several lensed AGNs \citep{Ding_2021}. As with the local AGNs, we recalibrate each sample using the \mgii-based \Mbh expression in \eqnref{mbh_L3000} or the \hbeta\xspace equation from \cite{Trakhtenbrot_2012}. 

D20 find that AGN-hosting galaxies at $z\sim 1-1.5$ have mildly overmassive SMBHs on average for a given stellar mass, compared to the local relation. Conversely, the WSLQ sample studied here appears consistent with the local relation within roughly $\pm1\sigma$, in line with \cite{Suh_2020} who find no evolution in the \massrelation relation since $z\sim 2.5$. Our updated \massrelation constraints in \figref{mass_plot_basic} do not rule out the null hypothesis of no evolution because of the intrinsic scatter in the local relation and uncertainties in the mass measurements for the previously studied AGNs.

\begin{figure}
    \centering
    \hspace*{-1.25cm}
    \includegraphics[width=270px]{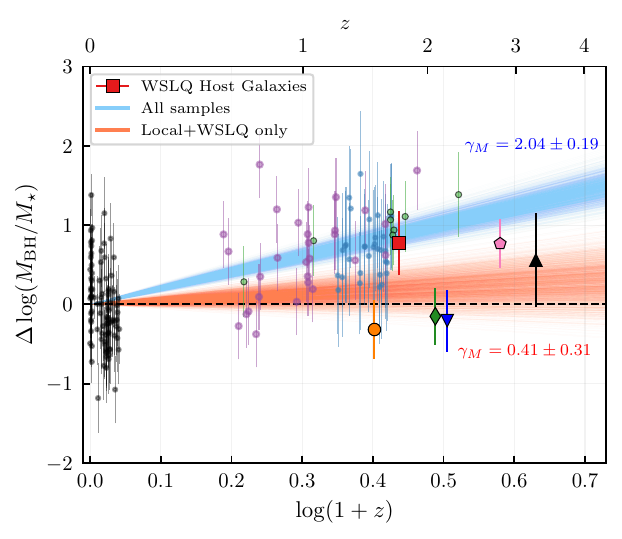}
    \caption{The offset from the measured local \massrelation relation plotted against redshift for all AGN samples, as computed in \eqnref{mass_offset}. The scattered datapoints are labeled identically as in \figref{mass_plot_basic}. The red and blue bands are random draws from the posterior of each fit, and illustrate the uncertainty. In particular, the blue band is the fit to all samples, and the red band is the fit to the WSLQs $+$ the local AGN `anchor' sample. The dotted line tracing $\Delta\log(\smbhmass/\stellarmass)$ = 0 corresponds to the local-relation best-fit in \figref{mass_plot_basic}. The region between $\pm 0.37$ on the $y$-axis is shaded in light grey to represent the intrinsic scatter in the local relation.}
    \label{fig:mass_evolution}
\end{figure}

Following recent examples in the literature \citep[e.g., D20;][]{Suh_2020, Li_2023}, and to directly compare \Mbh vs.\ \Mstar as a function of cosmic time, we compute an offset from the local relation model for all AGN host galaxies in \figref{mass_plot_basic}:
\begin{equation}
	\label{eqn:mass_offset}
	\Delta\log\left( \frac{\smbhmass}{\stellarmass} \right) \equiv \log{\left( {\frac{\smbhmass} {\Msun} } \right)} - \alpha_1 - \beta_1 \log{ \left( {\frac {\stellarmass} {10^{10} \Msun} } \right)}. 
\end{equation}
We visualize these offset values against redshift for the WSLQ systems and the other AGN samples in \figref{mass_evolution}. Then, we quantify the evolution in the \massrelation relation with redshift by fitting a power law to the offset:
\begin{equation}
	\label{eqn:mass_evolution}
	\Delta\log\left( \frac{\smbhmass}{\stellarmass} \right) = \gamma_M \log\left( 1 + z \right),
\end{equation}
where $\gamma_M$ is the slope.
Note that this form enforces the expected result of $\Delta\log\left( \smbhmass / \stellarmass \right) = 0$ at $z=0$. 

Using constraints from all samples shown in \figref{mass_plot_basic}, we find $\gamma_{M,\rm{lit}} = \litagnmodel$.
Treating the WSLQs separately from the literature samples, we find $\gamma_{,\rm{WSLQ}} = \wslqmodel$, suggesting a much smaller evolution - consistent with no evolution - for this subset. 
We quantify this discrepancy by drawing random samples from each fit's MCMC chain and computing the percentage of instances where $\gamma_{M,\rm{lit}} < \gamma_{M,\rm{WSLQ}}$. This test indicates a tension of $\sim 4.6$ significance between the WSLQ-based model and that with the literature samples. 
This tension must be the result of either differences in the analyses used to deduce either of the relevant mass measurements, or reflect an underlying difference in the samples analysed here. An obvious difference is redshift\textemdash four out of six WSLQs discussed here are at $z>2$, more distant than the vast majority of the literature samples. We explore both of these possibilities below.

\begin{figure}
    \centering
    \hspace*{-1cm}
    \includegraphics[width=270px]{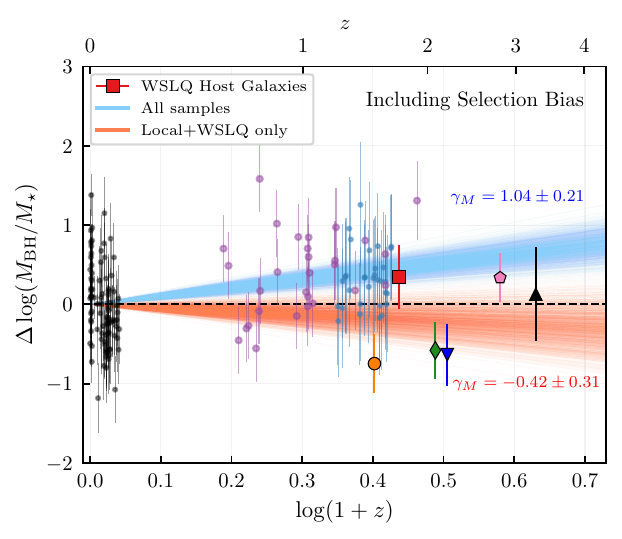}
    \caption{Identical to \figref{mass_evolution}, but after accounting for selection bias in the WSLQ sample and the other AGN samples at $z>0.5$. The two models are in a similar tension as before, where the WSLQs in our sample have smaller offset values on average. Note that the sample of lensed quasars from \cite{Ding_2021} is not included in this model because we do not have a selection function.}
    \label{fig:mass_evolsel}
\end{figure}

\subsection{Selection Effects}
\label{sec:cosmo_and_sel}

The intrinsic change in \dmbhmstar as a function of redshift has important physical implications for the co-evolution of AGNs and galaxies, and we expect cosmological probes of AGNs to reflect that evolution.
For example, a constant $\Delta\log\left( \smbhmass / \stellarmass \right)$  implies that one should see a similar trend with redshift between cosmic black hole and stellar mass densities, and between cosmic black hole accretion rate (BHAR) and SFR \citep{Mullaney_2012, Suh_2020}. 

However, we have to account for sample selection effects in order to interpret all samples in the context of AGN and galaxy populations. Following the method described later in \appref{sel_bias}, for each sample at redshift $z > 0.5$ including the WSLQs, we model the selection-induced offset \dmbhmstar we would expect to find if we assume that the \massrelation relation does \textit{not} evolve with redshift. After applying each offset as a bias correction to each corresponding sample, we re-fit the samples for \eqnref{mass_evolution}, one with the literature samples and the other with only the WSLQs. For the first model, we now find a power-law slope of $\gamma_{M,\rm{lit}} = \litagnmodelsel$, while for the WSLQ model we find $\gamma_{M,\rm{WSLQ}} = \wslqmodelsel$. These selection-corrected models produce a similar tension as before, of magnitude $\sim 3.9\sigma$, and we visualize the revised models and this difference in \figref{mass_evolsel}.

While our method for subtracting out the inferred selection biases slightly reduces the tension between the WSLQ sample and the literature samples, it is still subject to further potential uncertainties and biases. X-ray observations, widely used for selection of luminous AGNs \citep[including by D20;][]{Bennert_2011b, Schramm_2013}, give a (mostly) complete census of BH accretion except for in the most obscured systems \citep[e.g.,][]{Ricci_2015}. However, mildly more frequently (a factor of $\sim 2$), X-ray-selected AGNs reside in bluer, less massive, and more star-forming galaxies per traditional rest-UV-optical diagnostics \citep[e.g.,][]{Aird_2012}, implying that X-ray samples may not be complete in \Mstar even if they are in \Mbh. Furthermore, radio surveys provide a more complete view of less rapidly accreting AGNs in massive galaxies, which often live in galaxy groups and are more quiescent \citep[e.g.,][]{Best_2012}. A more complete study of selection effects in the \massrelation relation will require a joint analysis using both X-ray and radio selections to minimize bias in the host galaxy population. Given that the WSLQs are classical, unobscured broad-line quasars, we assume here that they are drawn from the same quasar population at $z\sim 2$ as found in X-ray surveys.

\subsection{Stellar Populations of the Hosts: SFR and \Mstar}
\label{sec:galaxyproperties}

\begin{figure*}
    \centering
    \includegraphics[width=15cm]{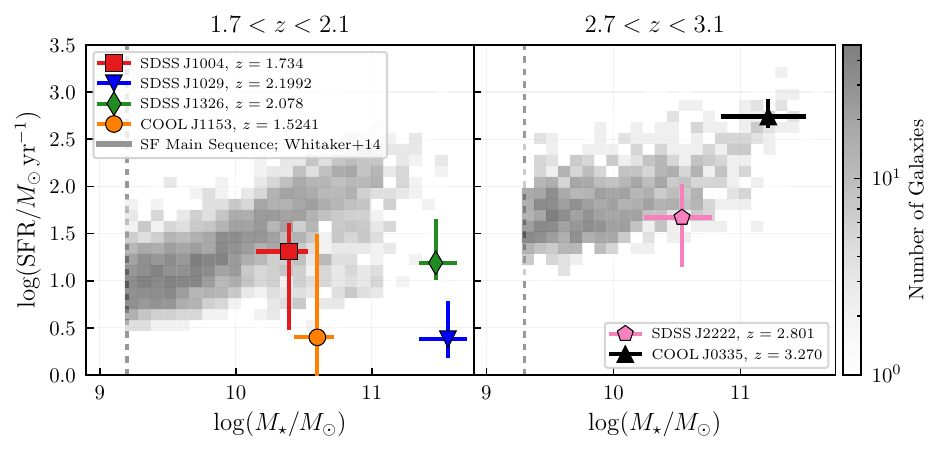}
    \caption{The WSLQ host galaxies plotted over the star-forming main sequence at respective redshifts, measured with \hst and \textit{Spitzer} photometry by \citet{Whitaker_2014}, from which the grey meshplots come. \twos and \fives fall below the main sequence, implying that they have already quenched or are in the process of quenching at the observed epoch. The SFR distribution for \eights is mostly below the main sequence but is still consistent within $+1-2\sigma$, leaving the interpretation more uncertain. \ones and \threes appear similarly consistent with lying on the main sequence but seem more likely to sit just below it. \sevens sits on the main sequence, appearing consistent with being a massive star-forming galaxy.}
    \label{fig:sf_mainseq}
\end{figure*}

For each WSLQ host galaxy, in order to evaluate their location on the star formation main sequence \citep[SFMS; e.g.,][]{Whitaker_2014, Speagle_2014, Leja_2022, Popesso_2023}, we compute their SFR and specific star formation rate (sSFR) from the models in \secref{spsmodeling}; these values are reported in \tabref{properties} and plotted in \figref{sf_mainseq}. 

Note that we expect the measured SFR posterior distributions for all SED fits in our sample to be overconfident (i.e., underestimating uncertainties), given the lack of constraints on dust-obscured star formation from mid- and far-IR observations, as well as the underlying challenge of isolating unobscured AGN light contribution from the emission from stellar populations.  

\begin{table*}
\hspace{-2.2cm}
\begin{tabular}{c c c c c c c c c} 
 \hline
 \multirow{2}{2em}{\centering Name} & \multirow{2}{3em}{\centering $\mu_{\rm{m}}$} & \multirow{2}{4em}{\centering Line}
 & \multirow{2}{4em}{\centering $\log(\lambda L_{\lambda})$} & \multirow{2}{5.5em}{\centering $\Delta V\, (\kms)$} 
 & \multirow{2}{6em}{\centering $\log(\smbhmass / \Msun)$} & \multirow{2}{6em}{\centering $\log(\stellarmass / \Msun)$} 
 & \multirow{2}{10em}{\centering $\log({\rm{SFR}}/ \Msunyr)$ } & \multirow{2}{7em}{\centering $\log({\rm{sSFR}}/ \yr^{-1})$} \\ [0.5ex] 
 \\
 \hline
\ones & $23.3\pm 7.2$ & \mgii & 44.00 & 2690 & $8.17\pm 0.33$ & $10.39_{-0.24}^{+0.14}$ & $1.31_{-0.83}^{+0.30}$ & $-8.94_{-0.41}^{+0.36}$ \\ 
\twos & $5.5 \pm 0.3$ & \mgii & 45.08 & 2554 & $8.35\pm 0.33$ & $11.56_{-0.21}^{+0.14}$ & $0.38_{-0.20}^{+0.40}$ & $-11.34_{-0.20}^{+0.43}$ \\
\threes & $12.0\pm 5.0$ & $\cdots$ & $\cdots$ & $\cdots$ & $8.31\pm 0.07$\footnote{from \cite{Williams_2021b}} & $10.54_{-0.30}^{+0.24}$ & $1.67_{-0.52}^{+0.36}$ & $-9.08_{-0.49}^{+0.65}$ \\
\fives & $3.0_{-1}^{+2}$ & \mgii & 44.52 & 3913 & $8.29\pm 0.33$ & $11.47_{-0.10}^{+0.15}$ & $1.19_{-0.18}^{+0.46}$ & $-10.45_{-0.16}^{+0.59}$ \\
\sevens & $5.3_{-1.8}^{+3.4}$ & \civ & 44.15 & 10017 & $8.79\pm 0.5$ & $11.22_{-0.37}^{+0.31}$ & $2.74_{-0.12}^{+0.19}$ & $-8.62_{-0.35}^{+0.43}$ \\
\eights & $5.2\pm 0.6$ & \mgii & 43.44 & 2774.6 & $7.29\pm 0.33$ & $10.60_{-0.17}^{+0.12}$ & $0.40_{-1.16}^{+1.09}$ & $-10.37_{-1.10}^{+1.15}$
\end{tabular}
\caption{\label{tab:properties}Source-plane quasar and host galaxy properties for the WSLQs studied in this work. Magnification values, used to transform measurements into the source plane are given in the second column, and they correspond to the green-circled images in \figref{wslq_images}. Columns from left to right: WSLQ name, magnification, broad emission line of choice for modeling \Mbh, quasar continuum luminosity in $\rm{erg\,s}^{-1}$, broad-line FWHM in $\kms$, SMBH mass, host stellar mass, host SFR, and host specific SFR. The sSFR is computed as ${\rm{SFR}} / M_{\star, \rm{tot}}$. Note that for \sevens, the FWHM reported has been corrected according to the measured centroid blueshift in \civ\xspace emission, as outlined in the text.
}
\end{table*}

Considering SFRs and stellar masses for each galaxy, we note that the host galaxies of \twos and \fives nominally fall below the `star-forming main sequence' at $z\sim 2$ \citep{Whitaker_2014}. We observe these galaxies as hosting a luminous quasar, and note the possibility that these galaxies have small instantaneous SFR, and hence maybe quenched. From a purely correlative standpoint, then, these galaxies appear consistent with a scenario in which feedback from their central quasars are linked to the regulation of star forming activity. 

We label the host galaxies of \ones, \threes, and \eights as `intermediate,' \ie, with some active star formation, where only further detailed spectroscopic followup may indicate hints of quenching or cessation of star formation.

For \threes, \cite{Bayliss_2017} find significant Lyman-$\alpha$ emission within $\sim 1 \kpc$ of the accretion disk, not associated with the quasar. The authors argue that this emission is physically consistent with circumnuclear clouds (i.e., within 1 kpc of the central SMBH) that absorb outflowing UV photons in their directions before they reach star-forming regions in the host galaxy. In other words, these clouds contribute to the `covering factor' of the quasar outflow \citep[e.g.,][in their Sec. 3.2]{Veilleux_2020}. Coupled with the more global presence of some star formation in the host galaxy, the physical properties of \threes correlate with a scenario in which circumnuclear gas might increase the timescale of regulatory quasar feedback processes.

\sevens is already a massive star-forming galaxy at $z = 3.270$, and while the SFR is more uncertain than it may appear, our SFR constraints and the detected quasar activity suggests some intriguing physical hypotheses. First of all, the incidence of both the massive quasar and the SFR is consistent with the presence of dense gas in both the circumnuclear region and in the ISM. Oddly, though, the quasar is not obscured by dust since we clearly see the rest-UV quasar continuum. IR observations of this galaxy in particular would be especially interesting; strong detections would indicate that the host galaxy is a `luminous IR galaxy,' and a strong far-IR detection could mean that \sevens is a rare `cold quasar' \citep{Kirkpatrick_2020}, a blue unobscured quasar within a starburst host galaxy.

\section{Systematics and Limitations}
\label{sec:systematics}

As already noted in \secref{results}, deducing evolutionary trends from a redshift-ordered set of incomplete cross-sectional samples is a potentially fraught process. Numerous potential systematics, some specific to individual samples, and others more broadly applicable, should be considered.
Specifically, strongly lensed sources may have biases related to lens modeling. All of the samples discussed here will have biases from the SED fitting treatment of heterogeneous data, influencing the resulting \Mstar and SFR measurements. 
The choice of comparison samples must also be considered. The potential systematics and the limitations they impose are considered below. 

\subsection{Uncertainties in Lensing Magnifications}
\label{sec:lensing_sys}

As lens modeling quantifies the degree to which lensed sources are magnified, accurate and precise lens models are crucial in order to transform observed properties of lensed sources to their intrinsic, unlensed values. To investigate lens modeling systematics, we recompute \Mbh and \Mstar for each WSLQ system (except for \threes) after perturbing the magnification $\mu_{\rm{m}}$ by each of $\pm 1\sigma$, a factor of $1/2$, and a factor of $2$. The motivation for this test is twofold. One idea is that the perturbed mass measurements give a more complete view of the statistical uncertainties in \Mbh and \Mstar. The other is that it provides a consistency check of our methodology. If both \Mstar and \Mbh depend on the lensing magnification, then perturbing the magnification should simply shift a lensed quasar's location in \massrelation parameter space in a direction roughly parallel to the local relation. 

\begin{figure}
    \centering
    \includegraphics[width=240px]{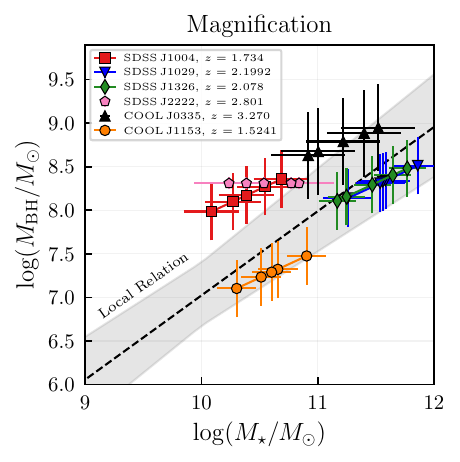}
    \caption{Similar to \figref{mass_plot_basic}, but now with $\mu_{\rm{m}}$-perturbed measurements of \Mbh and \Mstar. The magnifications shown in an array are $(\mu_{\rm{m}} / 2,\ \mu_{\rm{m}} - 1\sigma,\ \mu_{\rm{m}},\ \mu_{\rm{m}} + 1\sigma,\ 2\mu_{\rm{m}})$. Other samples shown in \figref{mass_plot_basic} are removed for the sake of visual clarity. Note that \threes is not perturbed in \Mbh because the reverberation mapping measurement is independent of magnification.}
    \label{fig:mass_plot_magn}
\end{figure}

As shown in \figref{mass_plot_magn}, each WSLQ's position in \massrelation parameter space relative to the local relation does not change significantly as we perturb the magnification. We therefore conclude the perturbed measurements do remain consistent with our results for the \massrelation relation in \secref{results}. Note, though, that an underestimated magnification would result in a slightly overestimated offset $\Delta\log\left( \smbhmass / \stellarmass \right)$ and vice versa. This trend stems from how the perturbation impacts \Mstar and \Mbh. While \Mstar scales linearly with $\mu_{\rm{m}}$, \Mbh does not, according to \eqnref{mbh_L3000}. Instead, $L_{3000}$ scales linearly with $\mu_{\rm{m}}$, and so \Mbh scales as $\mu^{0.62}$.

We must mention that a notable degree of magnification uncertainty does exist in the present archival data of WSLQs, even with the resolution of \hst. Such is especially the case for \fives, \sevens, and \eights, for which only ground-based imaging exists. A more careful analysis of each WSLQ's lens model, using better imaging data, is required to fully unpack the lens modeling systematics. This level of detail is beyond the scope of this work. At the same time, from our analysis, uncertainties in the magnification do not appear to significantly impact the results.

\subsubsection{Multiple Quasar Images}
\label{sec:imgs_sys}

In this work, we have so far analyzed one lensed image for each WSLQ system. If the methodology is robust and the measurements are accurate, then we ought to find the same results for a given quasar and host galaxy using different lensed images. To test that idea and further verify our methods, we measure and compare \Mbh, \Mstar, and other host galaxy properties between the three brightest images for two WSLQ systems: \ones and \eights.

For \ones and \eights, we replicate each target's respective analysis for each image as in \secref{analysis}, with some key differences to ensure self-consistency. In particular, our methods for obtaining image-plane photometry and for SED modeling are unchanged for each. For \ones we remagnify and then demagnify archival photometric measurements from \cite{Ross_2009}, who report host galaxy fluxes for its three brightest images. For \eights we perform SB modeling of images B and C using \galfit to extract host galaxy magnitudes, as in \secref{photometry}.

Regarding \Mbh and spectroscopy, for \ones, the eBOSS programs obtained a spectrum of image B as well as image A, but not of image C. To approximate a \mgii\xspace emission line of \ones's image C, we compute a flux ratio between images A and C in the $z$-band using the fluxes from the public DECaLS DR10 catalog. We then rescale the spectrum of image A by that ratio and repeat the spectroscopic analysis outlined in \secref{quasarspec}. We repeat this rescaling to estimate \Mbh for images B and C of \eights.

We have implicitly assumed that the difference between the magnification of the quasar and that of its host is negligible. As a simple test for the assumption that there is no difference, for \ones and \eights, we compute the average magnification of the host as follows. We draw a contour around the host flux in the image plane, compute the corresponding `undeflected' contour in the source plane, and then compute the image-to-source ratio of projected areas as the magnification. The statistical $1\sigma$ uncertainties are derived by repeating the process with 100 iterations of the lens model, randomly drawn from the MCMC posterior. Then, we rescale the photometry of each host galaxy image by its average magnification and recalculate \Mstar and other host galaxy properties. As a test for lensing and magnification systematics, note that our methods are not comprehensive, as other issues can easily arise such as magnification gradients, which may require greater attention to detail than this test gives \citep[e.g.,][]{Klein_2024}.

\begin{figure}
    \centering
    \includegraphics[width=240px]{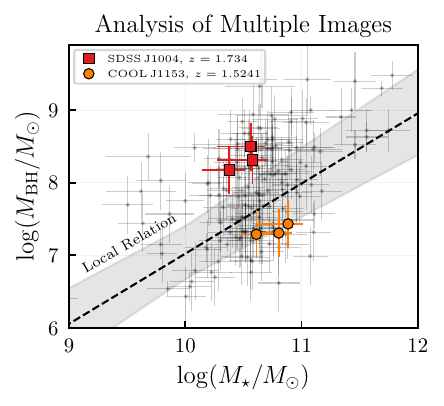}
    \caption{A nearly identical plot as \figref{mass_plot_magn}, instead comparing \Mbh and \Mstar of different quasar images for \ones, \sevens, and \eights. All of the samples from the literature are plotted as grey datapoints, and the local relation is the grey band.}
    \label{fig:mass_plot_imgmult}
\end{figure}

\Mstar and \Mbh derived from the multiple images of each system are shown in \figref{mass_plot_imgmult}. For both \ones and \eights, their three images appear consistent with one another within $\sim 1\sigma$, and the principal axis of dispersion appears to be visually similar to that traced by the perturbations in \figref{mass_plot_magn}. In addition, the three WSLQs remain consistent with what we find in \secref{massrelation}. This test for consistency between images implies that systematics related to lensing configurations, or differences between images, do not significantly influence our primary results in \secref{analysis}. Furthermore, this check also indicates that quasar variability is not an important issue either, since each image is captured at a different time along the light-curve due to time-delays \citep{Fohlmeister_2013, Dahle_2015, Munoz_2022, Napier_2023b}.

We list the different measurements of $\mu_{\rm{m}}$, \Mstar, and \Mbh for each image in \tabref{imgmulti}.

\begin{table}
\hspace*{-1.4cm}
\begin{tabular}{c c c c c} 
 \hline
 \multirow{2}{3em}{\centering Image} 
 & \multirow{2}{3em}{\centering $\mu_{\rm{m,QSO}}$} & \multirow{2}{3em}{\centering $\mu_{\rm{m,host}}$}
 & \multirow{2}{6em}{\centering $\log(\smbhmass / \Msun)$} & \multirow{2}{6em}{\centering $\log(\stellarmass / \Msun)$}  \\ [0.5ex] 
 \\
 \hline
 \ones-A & $23.3_{-7.2}^{+7.2}$ & $23.7_{-5.1}^{+6.0}$ & $8.17\pm 0.33$ & $10.38_{-0.24}^{+0.14}$ \\ 
 \ones-B & $13.8_{-3.8}^{+3.8}$ & $15.6_{-2.9}^{+3.6}$ & $8.50\pm 0.33$ & $10.56_{-0.29}^{+0.12}$ \\ 
 \ones-C & $10.7_{-1.3}^{+1.3}$ & $11.0_{-1.3}^{+1.3}$ & $8.32\pm 0.33$ & $10.58_{-0.30}^{+0.12}$ \\ 
 \eights-A & $5.2_{-0.6}^{+0.6}$ & $5.1_{-0.5}^{+0.7}$ & $7.29\pm 0.33$ & $10.61_{-0.17}^{+0.12}$ \\ 
 \eights-B & $3.9_{-1.2}^{+1.8}$ & $5.3_{-1.6}^{+2.1}$ & $7.43\pm 0.33$ & $10.88_{-0.27}^{+0.13}$ \\ 
 \eights-C & $5.3_{-0.9}^{+0.9}$ & $5.3_{-0.9}^{+0.9}$ & $7.31\pm 0.33$ & $10.80_{-0.23}^{+0.12}$ \\
\end{tabular}
\caption{\label{tab:imgmulti} Properties from analysis of multiple lensed images. The columns are as follows, from left to right: the quasar image, magnification as computed from a contour around quasar flux, magnification for a contour around host galaxy flux, black hole mass, and stellar mass. The magnifications of the host and quasar are consistent within $1\sigma$ for each quasar image.
}
\end{table}

\subsection{Stellar Population Synthesis Systematics}
\label{sec:sps_sys}

\subsubsection{Intrinsic \Mstar Offsets from SPS Models}

Constraining a galaxy's SFH is a nontrivial exercise. The associated parameter space is highly nonlinear, and nearly all SED models will need to make some simplifying assumptions in order to extract useful information \citep{Conroy_2013, Carnall_2019}. Various biases associated with SFH model choices will infiltrate into the results, and any robust study of galaxy SEDs and derived properties ought to account for these biases. To that end, we explore the impact of assuming the delayed-tau model on our \Mstar measurements. To do so, we take a random sample of 311 well-studied galaxies from the CANDELS field \citep{CANDELS_Santini_2015} at $0.5 < z < 5$. We fit each galaxy with a simple version of the SPS framework from \secref{spsmodeling}, with $M_{\star, \rm{tot}}$, $t_{\rm{age}}$, $\tau$, and $\tau_{\lambda,2}$ as free parameters, and without continuum or emission from interstellar nebulae or dust.

\begin{figure*}
    \centering
    \includegraphics[width=16cm]{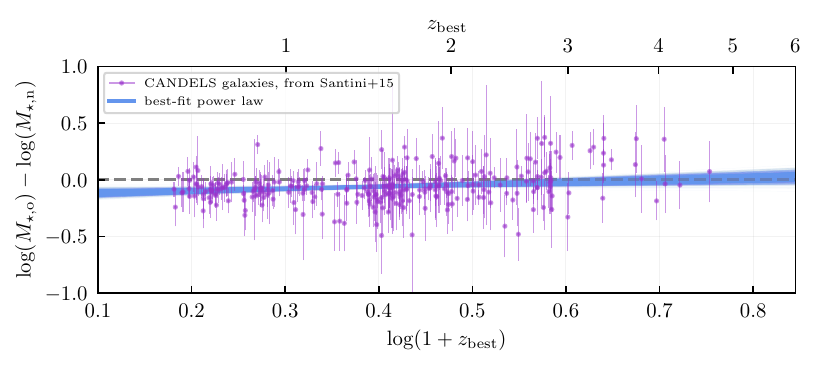}
    \caption{Residual between the \cite{CANDELS_Santini_2015} measurements $M_{\star , {\rm o} }$ and our measurements $M_{\star , {\rm n} }$ for CANDELS galaxies as a function of redshift. The blue band is a random sample of power-law fits computed using an MCMC scheme. We do not find significant offsets of concern, while noting that the SPS framework we employ may slightly overestimate \Mstar on average at lower redshift.}
    \label{fig:candels_sps}
\end{figure*}

We compare model results on CANDELS galaxies to the previous \Mstar measurements from \cite{CANDELS_Santini_2015} in order to probe intrinsic bias in the stellar mass constraints for WLSQ host galaxies. As shown in \figref{candels_sps}, we compute the difference between the catalog \Mstar values and the new values and measure the (residual) difference as a function of catalog redshift $z_{\rm{best}}$. Similarly to how we constrain the time-evolution of \dmbhmstar, we fit a power law (a logarithmic slope and intercept) with an MCMC code to characterize how the residual changes with redshift. Given the uncertainty shown on the power-law model and the distribution of datapoints at $z < 1.5$ slightly below 0, and specifically assuming that the measurements from \cite{CANDELS_Santini_2015} are accurate, the SPS modeling framework used here might overestimate \Mstar on average by $\sim 0.1$ dex at low redshift. More broadly, we suggest that uncertainties at this level should be taken as a bare minimum uncertainty \citep{Lower_2020} in the comparison between various literature quasar samples used here, each of which have their own SPS treatment.

\subsubsection{SPS Model Choices and Heterogeneity in Individual WSLQ Hosts}

In \secref{spsmodeling}, we adjusted the SPS model of a given WSLQ host galaxy based on the constraining power of available photometry.
To test for biases from this heterogeneity, we adopt a simpler model setup and a more complex, more flexible setup and apply each of these to the host galaxies of \twos and \fives. We choose these two hosts because they are least well-constrained by the available photometry with only three measurements each (\tabref{data}). 

Here, for dust attenuation, instead of \cite{Calzetti_2000}, we use the formalism of \cite{Kriek_2013} which uses two additional dust parameters. These are (a) the attenuation around young stellar populations and (b) the slope of the attenuation curve. The motivation for using this attenuation curve is that it is likely more suitable for distant galaxies. The simpler model keeps $M_{\star, \rm{tot}}$, $t_{\rm{age}}$, $\tau$, and attenuation around older stellar populations as free parameters and keeps stellar metallicity constant, while setting the two attenuations (older and younger) to be equal. The more complex model is similar to the model for \ones in \secref{spsmodeling}, while adding attenuation around young stars and the attenuation slope via the \cite{Kriek_2013} curve.
For \twos, the more complex model again contains a suspicious dusty solution that is likely too massive and star-forming at the same time; we again rule out this solution. 

\begin{figure*}
{
\centering
\begin{tabular}{c}
	\includegraphics[width=15cm]{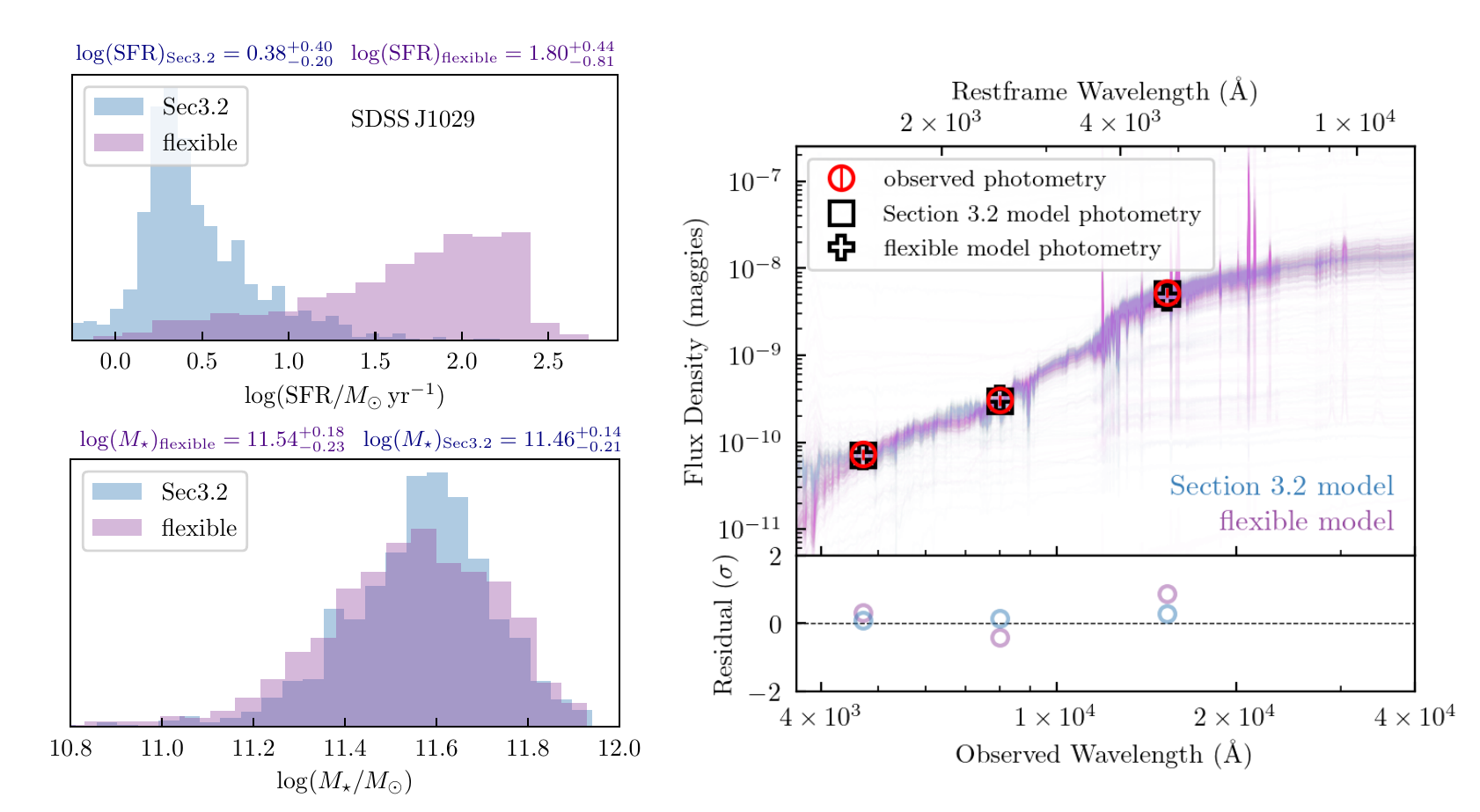} \\
	\includegraphics[width=15cm]{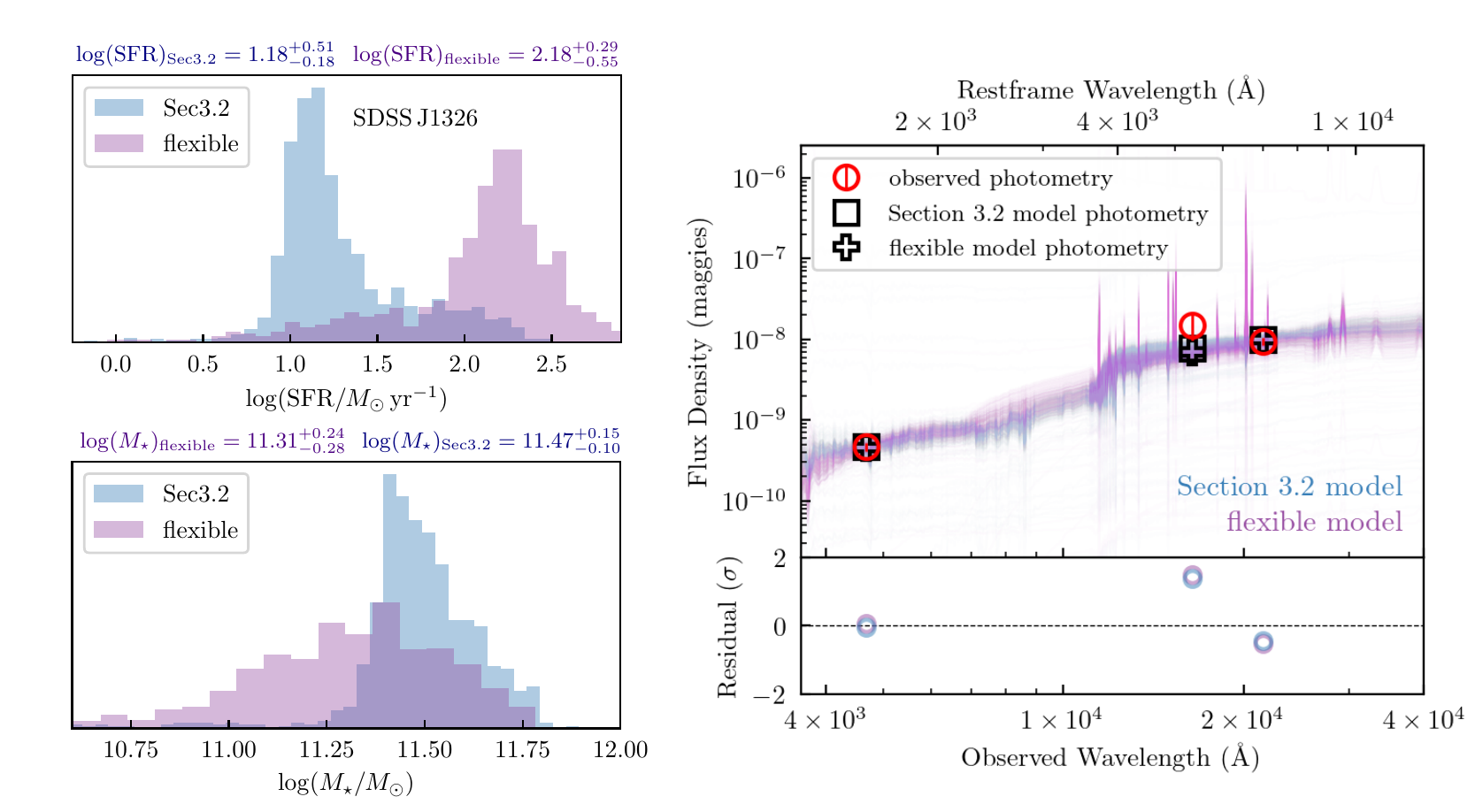} 
\end{tabular}
\caption{Comparison of output galaxy SEDs, \Mstar distributions, and SFR distributions between the models described in \secref{spsmodeling} and the more flexible models, for the host galaxies of \twos \textbf{(top)} and \fives \textbf{(bottom)}. Models from before are shown in light blue, while the more flexible uniform models are shown in purple/magenta. While the SED fits are similar from a statistical, least-squares perspective, the SFR constraints differ between the two models by $\gtrsim 1\sigma$ for both galaxies. }
    \label{fig:modelcompare}
}
\end{figure*}

We find that the host galaxy parameters of \twos and \fives are affected strongly by the model heterogeneity. The simpler model for \twos gives results which are consistent with the model from \secref{spsmodeling}, but the more flexible model gives a SFR posterior which is larger than before at a rate of $\sim 1.4\sigma$ even excluding the suspicious solution. We find a similar result with \fives, where the simpler model is more consistent but the flexible model gives a larger SFR by $\sim 1.2\sigma$. \figref{modelcompare} visualizes these differences in SFR measurements for both galaxies, alongside the stellar masses and SEDs for the corresponding models. 

Comparing the simple and flexible models directly, the inferred median stellar mass is less sensitive to the model choice, but the scatter in the distribution may change as it does for \fives. The SFR is much more sensitive, to the extent that it can change where a galaxy falls in the \Mstar\textendash SFR plot (star-forming, quiescent, or starburst).
While not consequential in every case, this outcome validates our expectation that SPS model choices influence the model results and the corresponding interpretations. Furthermore, it indicates that the actual physical uncertainties in SFR are larger than the statisitical ones given by the model, as we discussed in \secref{galaxyproperties}.
Follow-up IR observations and/or much deeper spectroscopy would be required to better constrain the host galaxy SEDs and their star formation histories.

On a more specific note about \twos, the host galaxy has been observed with \textit{Herschel Space Observatory}, as \cite{Stacey_2018} detected flux at each of 250, 350, and 500 $\mu$m (see their Table A1), indicating a presence of thermal emission from cold dust. In addition to a desire to preserve what homogeneity exists in the WSLQ sample, we elect not to include this far-IR photometry in the fiducial SPS model for \twos for two reasons. (1) If we add it and remove the upper limit on $\tau_{\lambda,2}$, we find that \Mstar and SFR are not significantly impacted. The corresponding dust constraint of $\tau_{\lambda,2} = 0.94_{-0.10}^{+0.20}$ does challenge the specific choice of $\tau_{\lambda,2} < 1$ from \secref{spsmodeling}, but it does not change our interpretations of \twos in the \massrelation relation or its stellar populations. The host galaxy is massive and it probably contains a small but notable amount of dust (given the faint far-IR detection), and its SFR is not well constrained.
(2) For sufficiently dusty clouds in a quasar host galaxy, the traditional assumption that cold dust emission is powered by star formation may break down \citep{McKinney_2021}. If the quasar produces at least some of the far-IR, then a model including \textit{Herschel} photometry would overestimate the host galaxy IR emission. We cannot confirm that the host galaxy is driving this far-IR emission in full, and including it here would make that assumption and therefore potentially bias the model posteriors.

The choice of stellar population model affects measurements both of WSLQs from this work and of AGNs from the literature. For example, D20 use simple stellar population templates of specific ages fit to the host galaxy photometry to constrain \Mstar.  They use the 21/32 host galaxies with robust photometric detections in two HST filters (either F140W or F125W, plus F814W) to choose a single template age for two redshift subsamples of $1.23<z<1.44$ and $1.44<z<1.67$. D20 then apply these same templates to the 11/32 systems which have no detection in the bluer filter (F814W), anchoring their results for these systems only on the NIR single-band photometry. Given the demonstrated range of host galaxy properties and colors\textemdash seen here and in extensive literature \citep[e.g.,][]{Dunlop_2003, Simmons_2012, Glikman_2015, Ding_2022a}\textemdash it seems equally plausible that this 1/3rd of their sample are instead actually older and redder host galaxies, whose masses will not be well measured by matching to a younger template.

Potentially, this measurement bias in \Mstar could account for some of the observed difference between the fits to the different samples in \secref{massrelation}; further data able to better constrain the stellar populations in these galaxies would be required to suppress this potential bias. Note that if we remove the D20 sample altogether, then we instead get a fit desribed by $\gamma_{M,\rm{lit}} = 1.22\pm 0.28$, which still differs from the WSLQ fit by $\sim 4.0\sigma$.

In addition, note that two of the known eight WSLQs were not included in the analysis presented here, as no detection of the host galaxy was apparent in the extant data. Notably, at the time of writing, the available data on both of these systems was only ground-based imaging of moderate seeing and depth, and so the lack of a host-galaxy detection may simply reflect the limits of available data. However, it is possible that these two systems have preferentially larger \dmbhmstar values, further lessening the few-sigma tension between the WSLQ results and other data.

\subsection{The Local Anchor and Self-Consistency}
\label{sec:local_anchor}

The choice of local sample used in `anchoring' higher redshift AGN samples can bias the resulting analysis of how galaxies and SMBHs co-evolve \citep[e.g., ][]{Bennert_2021, Li_2023}. To minimize this bias, a common solution is to use samples from the literature not only which are relatively complete (\appref{sel_bias}), but also to which the authors can easily apply their own methods (\ie, `self-consistency'). In other words, when choosing a local anchor, we ought to collate samples for which techniques as similar to ours as possible are used for \Mbh and \Mstar measurements. In line with this idea, we have used the local sample from \cite{Bennert_2021}, who use single-epoch spectroscopy for \Mbh and photometry and color information for \Mstar (albeit using a different code and set of templates for SED fitting). Furthermore, the \cite{Bennert_2021} local AGN sample is relatively diverse in terms of host galaxy morphology. However, taken alone the \cite{Bennert_2021} sample poorly constrains the slope of the local relation. Hence, following D20, in the above we supplemented this dataset with a nearby sample of massive quiescent galaxies from \cite{Haring_2004}, many of whose SMBHs are not actively accreting but which extends to higher \Mbh and thus better constrains the model of the local relation in \secref{massrelation}. Stellar masses for these systems are an aggregate of various detailed analyses from the literature, but the sample itself is almost exclusively early-type galaxies.
Given differences in methodology and host-galaxy populations between \cite{Haring_2004} and \cite{Bennert_2021}, and the higher-redshift comparisons, we may expect some systematic biases in stellar masses in our measurement of the \massrelation.

Self-consistency in \Mbh measurements is often more important and consequential, due to the larger uncertainty (sans reverberation mapping) and the wider variety of calibrations and different recipes \citep{Trakhtenbrot_2012, Ding_2020}. As such, for comparison to WSLQs, we have selected samples from other papers \citep[]{Bennert_2011b, Schramm_2013, Ding_2020, Ding_2021} which use single-epoch spectroscopic measurements of either \mgii\xspace or a Balmer line (\hbeta, H$\alpha$). We recalibrated \Mbh values from other papers in the analysis of the \massrelation relation where possible. 

However, even with self-consistency in \Mbh maximized, there are multiple \mgii-based calibrations of \Mbh in the literature and the choice of which to use impacts the results of the fits detailed in \secref{results}. If, rather than the formalism from \cite{Trakhtenbrot_2012}, we instead adopt the recipe given by \cite{Ding_2021} in their Section 2, we find a similar slope and intrinsic scatter but a higher normalization in the local relation by $\sim 0.3\,\rm{dex}$ (i.e., the local relation shifts upward). For the fit to the literature AGN samples, this change results in a best-fit offset \dmbhmstar which is consistent at $< 1\sigma$ with no \massrelation evolution, with $\gamma_{M,\rm{lit}} = 0.25\pm 0.27$ instead of $\gamma_{M,\rm{lit}} = \litagnmodelsel$ after accounting for selection effects. For the WSLQ-only fit, we find $\gamma_{M,\rm{WSLQ}} = -0.77\pm 0.33$ instead of $\gamma_{M,\rm{WSLQ}} = \wslqmodelsel$; i.e., a mild tension between the WSLQ measurement and that from other samples remains, but the overall picture is more consistent with no evolution.
Broadly, this is consistent with the models of \cite{Yang_2018}, \cite{Shankar_2020}, and \cite{Tanaka_2024}, whose results each span no evolution to mildly positive evolution in the \massrelation relation since $z\sim 2$. In other words, by that cosmic epoch the local \massrelation relation has mostly been established.

\section{Discussion}
\label{sec:discussion}

In formulating the black hole accretion rate (BHAR) as a function of galaxy stellar mass and redshift, with $\smbhmass / \stellarmass$ at $z=4$ as an `initial' condition, \cite{Yang_2018} infer a weak redshift evolution in the time-averaged ratio between BHAR and SFR from $z\sim 3$ to $z\sim 0.8$. This result corresponds to weak evolution (if any) in the \massrelation relation over that time. 
As both the BHAR and SFR densities are much lower at $z \lesssim 0.8$ than at $z\gtrsim 2$, the authors assume that both are essentially negligible after $z \sim 0.8$, and therefore the local \massrelation relation has nearly been established by $z\sim 2$.

The full selection-corrected fit (including WSLQs, samples from the literature, and considering selection biases) discussed in detail above exhibits at most only mild evolution in $\smbhmass / \stellarmass$ since $z\sim 2$. This interpretation is more of an upper limit, as the fitted index of the power-law evolution in redshift $\gamma_{M,\rm{lit}} = \litagnmodelsel$ gives offsets of similar magnitude to the intrinsic scatter of the local relation at $z=0$. 
Therefore, the fit to the literature data  ($\gamma_{M,\rm{lit}}$) appears consistent with \cite{Yang_2018}.
The WSLQ-only model ($\gamma_{M,\rm{WSLQ}}$) is consistent with no evolution, at $\sim 1.2\sigma$ below \dmbhmstar$=0$ before including intrinsic scatter $\sigma_{\rm{int}}$.

Moreover, we remind the reader that the redshift power-law evolution explored in prior papers and used here is almost certainly not an adequate description of the complex evolution of AGNs and their host galaxies over the entirety of cosmic time, and that the construction of an evolutionary sequence from time-ordered, cross-sectional samples selected and studied in various ways is a challenging process, at best. 

Recently, early observational works with \jwst have found a much larger \dmbhmstar in the early universe ($z \gtrsim 5$) \citep[e.g.][]{Maiolino_2023, Pacucci_2023, Stone_2023}, though whether that detection is intrinsic or merely a product of selection bias is not yet clear \citep{Li_2024}. If real, this larger \dmbhmstar at early times would support more rapid cosmic SMBH growth at high redshift and/or `heavy' SMBH seeding \citep[e.g.,][]{Schneider_2023}, followed by starburst activity such that \Mstar `catches up' by a later time. In that context, the small time evolution in \dmbhmstar since $z\sim 2$ shown by the literature samples in \figref{mass_evolsel} would represent an intermediate point between very early and late times, by which \Mstar has mostly but not entirely caught up.

Both a larger average offset at higher redshift $z \gtrsim 5$ and that of a negligible offset at that time remain open possibilities.
The WSLQ-only fit on its own is more consistent with a smaller offset (i.e., less rapid accretion and/or a different SMBH seeding scenario) at early times, and informs a redshift interval ($2<z<3$) where there are at present few other constraints. Interestingly, assuming our measurements are accurate and that the fit to the literature data is closer to the `true' evolution, the WSLQs' smaller \dmbhmstar values imply that they would fall on either the negative tail of the BHAR distribution function or the positive tail of the SFR distribution function \citep[each time-averaged;][]{Yang_2018} for galaxies hosting unobscured broad-line quasars. 

Regardless, these results clearly highlight the need for more detailed, \textit{multi-wavelength} reconstructions of AGN-fueled outflows, star formation histories, and conditions of the interstellar media in tandem and in greater detail at higher redshift. Spatially cross-correlating physical imprints of AGN activity with interstellar gas physics and star formation could illuminate the role(s) of AGNs in galaxy evolution and star formation in much more detail. Considering the level of precision in our measurements of WSLQs using relatively unexceptional (\ie, not groundbreaking) data quality, we emphasize that WSLQs and other highly magnified AGNs are among the most intriguing targets for these types of detailed study at any redshift they can be discovered \citep[e.g.,][]{Furtak_2023}.

\section{Summary}
\label{sec:summary}

With the available archival data and new observations obtained by the COOL-LAMPS collaboration, we use a sample of six unobscured broad-line quasars strongly lensed by massive galaxy clusters in order to study the \massrelation relation and the co-evolution of SMBHs and galaxies from $z\sim 3$ to $z = 0$. These `wide-separation lensed quasars' are among the most highly magnified multiply-imaged quasars known.
More specifically, we constrain quasar host galaxy properties (stellar mass and star formation rates, primarily) via the stellar population synthesis code \prospector. We constrain the quasar black hole masses by modeling the flux profile of a chosen rest-UV broad emission line, observed with single-epoch spectroscopy. These physical properties provide constraints on the time evolution of the \massrelation relation. In cases where the analysis is possible, we analyze multiple images of a given quasar, and demonstrate good agreement in results from these different images. 

We model potential evolution in the \massrelation relation from $z \sim 0$ as a power law with a normalization coefficient and a parameter for intrinsic scatter. We consider two fits describing the time evolution in \massrelation, parametrized by \dmbhmstar in \eqnref{mass_offset}: one fit to a set of other mostly unlensed samples at different\textemdash and mostly lower\textemdash redshifts, and one fit to only the WSLQs and a local AGN sample. 

The primary conclusions in this paper are as follows:
\begin{itemize}
    \item Highly magnified AGNs are among the most intriguing individual systems for future detailed study of AGN outflows, feedback, and the physical relation(s) therein to galaxy evolution. Even when relying on ground-based imaging, we have in some cases confidently detected each WSLQ's host galaxy flux and placed strong constraints on the stellar mass. This outcome is remarkable and possibly unique to WSLQs among luminous broad-line quasars; at cosmological distances, much higher resolution data from over-subscribed observatories such as \hst, ALMA, or \jwst is usually required to at all constrain host galaxy masses, luminosities, and other properties \citep[e.g.,][]{Peng_2006, Li_2023, Tanaka_2024}. Pairing such observations with the advantages of cluster-scale lensing could give further powerful constraints on \agnhost co-evolution and feedback at high redshift, particularly with the expanded samples expected from the Rubin Observatory \citep{Napier_2023b}. 
    \item The WSLQs are consistent with the local \massrelation relation and therefore with a scenario in which $\smbhmass/\stellarmass$ is relatively constant since $z\sim 3$. This result is contrasts with a fit to some AGN samples from the literature, with which the fit to the WSLQ sample differs by $4\sigma$.
    \item The WSLQ hosts are a mixture of star-forming and quiescent galaxy candidates. \twos and \fives are quiescent galaxy candidates, as some models of their stellar populations give SFRs which fall below the star-forming `main sequence.' \ones, \threes, and \eights have less constrained SFRs, given the existing data fit with a wide range of SPS models. \sevens is most consistent with a massive star-forming galaxy, despite the quasar's unobscured nature\textemdash we hypothesize that if the host galaxy is strongly detected in the far-IR, then it may be a rare `cold quasar' in addition to an absorption-line quasar \citep{Kirkpatrick_2020, Napier_2023a}. Currently available data is insufficient to robustly confirm or rule out scenarios for these host galaxy stellar populations. As such we avoid definitive statements on star formation histories, for which stronger constraints would require follow-up multi-wavelength observations. Whether individual WSLQ host galaxies are star-forming or quiescent, they represent intriguing future case studies for AGN feedback, as different types of stellar populations may represent various co-evolutionary stages between SMBHs and galaxies \citep{Hopkins_2006}.
    \item We investigate several sources of systematic uncertainty and make a number of simple tests, to verify or further explore the results for WSLQ host galaxy properties and the \massrelation relation. Potentially significant systematic issues include: selection bias (\secref{cosmo_and_sel}, \appref{sel_bias}), the SPS modeling framework (\secref{sps_sys}), and the choice of local anchor or of \Mbh calibration (\secref{local_anchor}). Additionally we explore the impact of uncertain magnifications and of the choice between different lensed images (\secref{lensing_sys}) The selection effects and the local anchor are both important, but none of the other choices significantly affect our interpretations. 
\end{itemize}

Upcoming wide-area surveys (\textit{Euclid}, LSST, \textit{Roman}) will drastically increase the discovery space and should grow the sample size of strongly lensed AGNs, including WSLQs. Additionally, the recent developments in IR observations make possible a number of strong potential improvements to our results and directions for additional study. \jwst imaging at NIR and mid-IR wavelengths would significantly improve the lens modeling \citep{Napier_2023b}, but it would also greatly facilitate bulge\textendash disk decomposition to study the $\smbhmass$\textendash $M_{\star,{\rm{bulge}}}$ relation. We can employ IR spectroscopy for a number of purposes, including: (1) improvements in the \Mbh constraints from access to the brightest Balmer lines\,\textemdash\, which correlate best with \Mbh\,\textemdash\, at all redshifts;  
(2) mapping accretion-powered outflows and their physical connection to star-forming regions in the host galaxy to unprecedented precision \citep{Cresci_2023}; and (3) constraining host galaxy kinematics with far-IR observations and studying the relation between \Mbh and the stellar velocity dispersion beyond redshift $z\sim1$. From both photometric and spectroscopic data, SPS modeling of the host galaxy SEDs would enormously improve in two ways; one in the stronger constraints on star formation histories given by broader wavelength coverage and the incorporation of spectroscopy; and two in the newfound ability to model the SEDs pixel-by-pixel \citep[e.g.,][]{Ding_2022a} to measure projected distributions of \Mstar and SFR.

\section*{Acknowledgments}
\label{ack}

This work was supported by The College Undergraduate program and the Department of Astronomy and Astrophysics at the University of Chicago. APC acknowledges funding from the College Center for Research and Fellowships at the University of Chicago.

We would like to express gratitude towards the staff at the 6.5m Magellan Telescopes at the Las Campanas Observatory, Chile for their valuable labor, and in particular for their care and efforts during a global pandemic.

This paper is partially based on data gathered with the 6.5m Magellan Telescopes located at Las Campanas Observatory, Chile. Magellan observing time for this program was granted by the time allocation committees of the University of Chicago and the University of Michigan.

This paper is partially based on observations made with the Nordic Optical Telescope, owned in collaboration by the University of Turku and Aarhus University, and operated jointly by Aarhus University, the University of Turku and the University of Oslo, representing Denmark, Finland and Norway, the University of Iceland and Stockholm University at the Observatorio del Roque de los Muchachos, La Palma, Spain, of the Instituto de Astrof\'isica  de Canarias.
The data presented here were obtained in part with ALFOSC, which is provided by the Instituto de Astrof\'isica de Andaluc\'ia (IAA) under a joint agreement with the University of Copenhagen and NOT.

This paper is partially based on observations made with the NASA/ESA Hubble Space Telescope, obtained from the Multimission Archive at the Space Telescope Science Institute (MAST) at the Space Telescope Science Institute, which is operated by the Association of Universities for Research in Astronomy, Incorporated, under NASA contract NAS 5-26555. These archival observations are associated with programs GO-10509, GO-9744, GO-10793, GO-12195, and GO-13337.

The Legacy Surveys consist of three individual and complementary projects: the Dark Energy Camera Legacy Survey (DECaLS; Proposal ID 2014B-0404; PIs: David Schlegel and Arjun Dey), the Beijing-Arizona Sky Survey (BASS; NOAO Prop. ID 2015A-0801; PIs: Zhou Xu and Xiaohui Fan), and the Mayall z-band Legacy Survey (MzLS; Prop. ID 2016A-0453; PI: Arjun Dey). DECaLS, BASS and MzLS together include data obtained, respectively, at the Blanco telescope, Cerro Tololo Inter-American Observatory, NSF’s NOIRLab; the Bok telescope, Steward Observatory, University of Arizona; and the Mayall telescope, Kitt Peak National Observatory, NOIRLab. The Legacy Surveys project is honored to be permitted to conduct astronomical research on Iolkam Du’ag (Kitt Peak), a mountain with particular significance to the Tohono O’odham Nation.

NOIRLab is operated by the Association of Universities for Research in Astronomy (AURA) under a cooperative agreement with the National Science Foundation. LBNL is managed by the Regents of the University of California under contract to the U.S. Department of Energy.

This project used data obtained with the Dark Energy Camera (DECam), which was constructed by the Dark Energy Survey (DES) collaboration. Funding for the DES Projects has been provided by the US Department of Energy, the US National Science Foundation, the Ministry of Science and Education of Spain, the Science and Technology Facilities Council of the United Kingdom, the Higher Education Funding Council for England, the National Center for Supercomputing Applications at the University of Illinois at Urbana-Champaign, the Kavli Institute for Cosmological Physics at the University of Chicago, Center for Cosmology and Astro-Particle Physics at the Ohio State University, the Mitchell Institute for Fundamental Physics and Astronomy at Texas A\&M University, Financiadora de Estudos e Projetos, Fundação Carlos Chagas Filho de Amparo à Pesquisa do Estado do Rio de Janeiro, Conselho Nacional de Desenvolvimento Cient\'ifico e Tecnol\'ogico and the Ministério da Ciência, Tecnologia e Inovação, the Deutsche Forschungsgemeinschaft and the Collaborating Institutions in the Dark Energy Survey. 
The Collaborating Institutions are Argonne National Laboratory, the University of California at Santa Cruz, the University of Cambridge, Centro de Investigaciones Enérgeticas, Medioambientales y Tecnológicas–Madrid, the University of Chicago, University College London, the DES-Brazil Consortium, the University of Edinburgh, the Eidgenössische Technische Hochschule (ETH) Zürich, Fermi National Accelerator Laboratory, the University of Illinois at Urbana-Champaign, the Institut de Ciències de l’Espai (IEEC/CSIC), the Institut de Física d’Altes Energies, Lawrence Berkeley National Laboratory, the Ludwig-Maximilians Universität München and the associated Excellence Cluster Universe, the University of Michigan, NSF’s NOIRLab, the University of Nottingham, the Ohio State University, the OzDES Membership Consortium, the University of Pennsylvania, the University of Portsmouth, SLAC National Accelerator Laboratory, Stanford University, the University of Sussex, and Texas A\&M University.

BASS is a key project of the Telescope Access Program (TAP), which has been funded by the National Astronomical Observatories of China, the Chinese Academy of Sciences (the Strategic Priority Research Program “The Emergence of Cosmological Structures” Grant \# XDB09000000), and the Special Fund for Astronomy from the Ministry of Finance. The BASS is also supported by the External Cooperation Program of Chinese Academy of Sciences (Grant \# 114A11KYSB20160057), and Chinese National Natural Science Foundation (Grant \# 12120101003, \# 11433005).

The Legacy Survey team makes use of data products from the Near-Earth Object Wide-field Infrared Survey Explorer (NEOWISE), which is a project of the Jet Propulsion Laboratory/California Institute of Technology. NEOWISE is funded by the National Aeronautics and Space Administration.

The Legacy Surveys imaging of the DESI footprint is supported by the Director, Office of Science, Office of High Energy Physics of the U.S. Department of Energy under Contract No. DE-AC02- 05CH1123, by the National Energy Research Scientific Computing Center, a DOE Office of Science User Facility under the same contract; and by the U.S. National Science Foundation, Division of Astronomical Sciences under Contract No. AST-0950945 to NOIRLab.

Funding for the Sloan Digital Sky Survey IV has been provided by the Alfred P. Sloan Foundation, the U.S. Department of Energy Office of Science, and the Participating Institutions. 

SDSS-IV acknowledges support and resources from the Center for High Performance Computing  at the University of Utah. The SDSS website is www.sdss4.org.

SDSS-IV is managed by the Astrophysical Research Consortium for the Participating Institutions of the SDSS Collaboration including the Brazilian Participation Group, the Carnegie Institution for Science, Carnegie Mellon University, Center for Astrophysics $\vert$ Harvard \& Smithsonian, the Chilean Participation Group, the French Participation Group, Instituto de Astrof\'isica de Canarias, The Johns Hopkins University, Kavli Institute for the Physics and Mathematics of the Universe (IPMU) / University of Tokyo, the Korean Participation Group, Lawrence Berkeley National Laboratory, Leibniz Institut f\"ur Astrophysik Potsdam (AIP),  Max-Planck-Institut f\"ur Astronomie (MPIA Heidelberg), Max-Planck-Institut f\"ur Astrophysik (MPA Garching), Max-Planck-Institut f\"ur Extraterrestrische Physik (MPE), National Astronomical Observatories of China, New Mexico State University, New York University, University of Notre Dame, Observat\'orio Nacional / MCTI, The Ohio State University, Pennsylvania State University, Shanghai Astronomical Observatory, United Kingdom Participation Group, Universidad Nacional Aut\'onoma de M\'exico, University of Arizona, University of Colorado Boulder, University of Oxford, University of Portsmouth, University of Utah, University of Virginia, University of Washington, University of Wisconsin, Vanderbilt University, and Yale University.

%-------------------------------------------------------------------------------

\facilities{Magellan Telescopes 6.5m (Baade/FourStar), Nordic Optical Telescope (ALFOSC, NOTCAM), Keck Observatory (LRIS), \hst}

\software{\numpy \citep{numpy_paper},
          \matplotlib \citep{matplotlib_paper}, 
          \astropy \citep{astropy:2013, astropy:2018}, 
          \emcee \citep{emcee_paper}
          \sextractor \citep{sextractor_paper}, 
          \galfit \citep{GALFIT_2002, GALFIT_2010},
          \prospector \citep{prospector_paper},
          \dynesty \citep{dynesty_2020},
          \pyspeckit \citep{pyspeckit_ascl}
          \dsnine \citep{ds9_2003},
          \lenstool \citep{lenstool_2007}
          }

\bibliography{main.bib}

\appendix
\restartappendixnumbering
\section{Forward-Modeling of Selection Effects}
\label{app:sel_bias}

Here we outline how we estimate selection bias \citep{Lauer_2007}, the method for which is based on that of \cite{Schulze_2011, Schulze_2014} (in particular, the `flux-limited' variant introduced in their Section 3). Their formalism allows for computation of an observed bivariate distribution function in \Mstar and \Mbh, given: a local \massrelation relation, a galaxy stellar mass function (GSMF), an AGN duty cycle (the fraction of SMBHs that are actively accreting) as a function of \Mbh, an Eddington ratio distribution function (ERDF), and a observational selection function. In our adaptation of this framework, for a given AGN sample we adopt a characteristic redshift (approximately the median) and forward-model a population of broad-line AGNs. For the mock dataset, we compute an average offset \dmbhmstar, which corresponds to the offset we would expect to measure assuming no intrinsic evolution in the \massrelation relation. The advantage of this technique is that it assumes little about available observational data. Even though we obviously do not have data for a large population of magnified quasars, this method can uncover information about the intrinsic selection effects in the WSLQ sample regardless. The input redshifts, selection functions, and predicted offsets for each AGN sample are listed in \tabref{selfunctions}. Note that these predicted offsets are what we use in our recalculation of cosmic evolution in the \massrelation relation from \secref{cosmo_and_sel}.

\begin{table*}
\centering
\begin{tabular}{c c c c} 
 \hline
 \multirow{2}{8em}{\centering AGN Sample} 
 & \multirow{2}{5em}{\centering Redshift} & \multirow{2}{14em}{\centering Selection Function}
 & \multirow{2}{6em}{\centering Predicted Offset} \\ [0.3ex] 
 \\
 \hline
 \cite{Schramm_2013} & 0.8 & \multirow{2}{14em}{$(43.7 < \log L_{\rm{bol}} < 45.7)\ and$ $(7.0 < \log\smbhmass < 9.4)$} & $+0.18$ \\ [3.5ex]
 \cite{Bennert_2011b} & 1.2 & \multirow{2}{14em}{$(43.5 < \log L_{X} < 44.5)\ and$ $(\log\smbhmass > 7.8)$} & $+0.38$ \\ [3.5ex]
 \cite{Ding_2020} & 1.5 & \centering \multirow{2}{14em}{$(7.7 < \log \smbhmass < 8.8)\ and$ $(45 < \log L_{\rm{bol}} < 46.2)\ and$ $(-2 < \log \lambda_{\rm{Edd}} < 0.5)$} & $+0.39$ \\ [5ex]
 WSLQs (this work) & 2.1 & \centering \multirow{2}{14em}{$(3 < \mu_{\rm{m}} < 30.5)\ and$ $(18.4 < r_{\rm{img}} < 22.12)$} & $+0.43$ \\ [3ex]
\end{tabular}
\caption{\label{tab:selfunctions} Inputs and output offset for modeling selection bias of each sample at redshift $z>0.5$, given the GSMF and ERDF each as a function of redshift. From left to right, the columns are: the paper of the given AGN sample, the characteristic (median) redshift of the AGN sample, the selection function for each sample as outlined in the corresponding paper, and the expected offset assuming no cosmic evolution in the \massrelation relation. Note that for the literature samples \citep[][D20]{Schramm_2013, Bennert_2011b}, we shift the \Mbh cuts slightly to reflect the recalibration to the framework from \cite{Trakhtenbrot_2012} (see our \secref{measure_mbh}). For the \cite{Bennert_2011b} sample, $L_X$ is the X-ray luminosity over the $0.5-8$ keV range, and we calculate this luminosity using a bolometric correction from \cite{Lusso_2012}.
}
\end{table*}

\begin{figure*}
    \centering
    \includegraphics[width=400px]{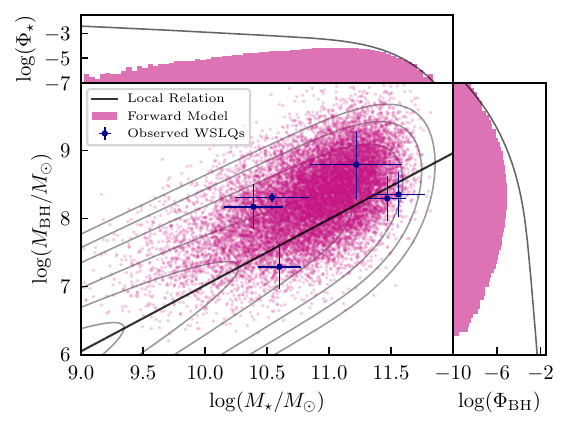}
    \caption{Forward modeling of a population of highly magnified quasars, akin to WSLQs, in \massrelation parameter space. \textbf{(Center)} The magenta scatter plot are the mock quasars, while the navy datapoints are the observed WSLQs studied in this paper. The black log-linear line is the measured local relation with slope $0.97$ and intercept $7.02$, used as an input in the forward modeling framework. The grey contours represent the bivariate distribution function in \massrelation parameter space, which we compute while accounting for the single-epoch \Mbh uncertainty\textemdash see \href{https://www.aanda.org/articles/aa/full_html/2011/11/aa17564-11/aa17564-11.html\#S5}{Section 3.4} and Eq. 24 in \cite{Schulze_2011}. \textbf{(Top)} The projection of bivariate distributions to the galaxy stellar mass, i.e., the GSMF in grey (solid line) and the `observed' \Mstar distribution of mock quasar host galaxies in magenta. The mock observed distribution is rescaled to emphasize the similarities and differences in shape with the GSMF. \textbf{(Right)} The projection in black hole mass, i.e., the black hole mass function in grey (solid line) and the `observed' \Mbh distribution of mock quasars in magenta (again rescaled).}
    \label{fig:sel_bias}
\end{figure*}

\begin{figure}
    \centering
    \includegraphics[width=231px]{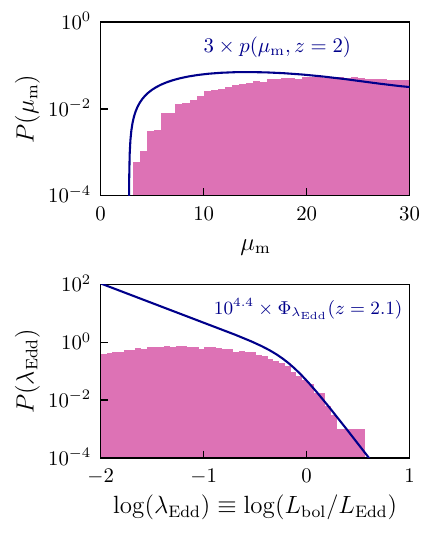}
    \caption{The input distributions for lensing magnification $\mu_{\rm{m}}$ \textbf{(top)} and Eddington ratio $\lambda_{\rm{Edd}}$ \textbf{(bottom)} used in the forward modeling framework, plotted against the respective histograms for the mock sample of magnified quasars. Each of the input distributions are rescaled for the sake of comparing the shapes of intrinsic vs. `observed' distributions, as in \figref{sel_bias}. Note that $p(\mu_{\rm{m}})$ is a \textit{probability} distribution while $\Phi_{\lambda_{\rm{Edd}}}$ is a physical distribution function like the GSMF.}
    \label{fig:input_dist}
\end{figure}

We use the following cosmological assumptions: (1) our measurement of the local relation as in \eqnref{mass_powerlaw}, with $\alpha_1 = 7.02\pm 0.38$, $\beta_1 = 0.97\pm 0.11$ and intrinsic scatter $\sigma_{\rm{int}} = 0.37\pm 0.05$; (2) the GSMF from \cite{Weaver_2023}, who fit a double Schechter function to each redshift bin; (3) the ERDF from \cite{Kelly_2013}, for which we fit a broken power-law model to each redshift bin; and (4) a constant AGN duty cycle of $p_{\rm{ac}} = 0.1$.
The GSMF is shown in the upper panel of \figref{sel_bias}, while the ERDF is shown (albeit rescaled) in \figref{input_dist}. The lattermost assumption of a constant duty cycle is a simplification that is fairly consistent with observational data of optically luminous broad-line quasars at $z \sim 2$ \citep[e.g.,][]{Kelly_2013}. We qualify that this approximation may begin to break down prior to $z = 2$, where the most massive SMBHs appear more likely to be AGNs than less massive ones \citep{Schulze_2015}.

We carry out the forward modeling as follows. A random array of mock galaxies with stellar masses \Mstar is drawn from the GSMF at the appropriate redshift. Black hole masses \Mbh are computed via the local \massrelation relation:
\begin{equation}
\begin{aligned}
	\label{eqn:mbh_forward}
	\log{\left( {\frac{\smbhmass} {\Msun} } \right)} = 7.02 + 0.97 \log{\left( {\frac{\stellarmass} {10^{10} \Msun} } \right)} \\
	+ n(\sigma_{\rm{int}}) + n(\sigma_{\rm{se}}),
\end{aligned}
\end{equation}
where we add two random Gaussian numbers: one representing our estimate for the intrinsic scatter $\sigma_{\rm{int}} = 0.37$ in \massrelation, and the other representing the scatter $\sigma_{\rm{se}} = 0.33$ in the single-epoch \Mbh measurements \citep{Trakhtenbrot_2012}. Then, as we have assumed $p_{\rm{ac}} = 0.1$, a random 10\% of the galaxies are assumed to host broad-line quasars. For these mock quasars, we randomly generate Eddington ratios in the range $\log\lambda_{\rm{Edd}} \in [-2, 1]$, weighted by the ERDF. At this point, bolometric luminosities are given as in \cite{Schulze_2011} by:
\begin{equation}
	\label{eqn:lbol_forward}
	\log L_{\rm{bol}} = \log\lambda_{\rm{Edd}} + \log{\smbhmass} + 38.1.
\end{equation} 

Finally, we select the mock population according to the selection function shown in \tabref{selfunctions}. Regarding this step, we note that the WSLQ selection function is more complex than those of the examples from \cite{Schulze_2011} because we select WSLQs using image-plane, magnified quasar photometry \citep{Martinez_2023, Napier_2023a, Kisare_2024}. Therefore, the WSLQ selection function is dependent on both observed flux and on lensing magnification, as opposed to simply on observed flux as in \cite{Merloni_2010} or on intrinsic parameters (\Mbh, Eddington ratio, AGN X-ray or bolometric luminosity) as in the papers for the other samples we employ \citep[][D20]{Bennert_2011b, Schramm_2013}. 

As an input for lensing magnification for the WSLQ sample, we apply a smoothing kernel to a $z=2$ magnification distribution function, computed numerically by \cite{Takahashi_2011} via ray-tracing simulations, such as to avoid overfitting and better approximate a `shape' of the distribution.\footnote{We remove the contribution from what \cite{Takahashi_2011} refer to as `Type I' lensing. In their terminology, `Types II and III' correspond to what is frequently called strong lensing. We direct readers toward their paper for further details.} This smoothed distribution function is visualized in the top panel of \figref{input_dist}.
For the image-plane flux, we convert $L_{\rm{bol}}$ to $r$-band magnitudes using a $K$-correction from \cite{Croom_2009} of $K(z) = -2.5(1+\alpha_\nu)\log(1+z)$ (where $\alpha_\nu$ is the spectral index) and the composite optically luminous quasar SED from \cite{Richards_2006}, which has a spectral index of $\alpha_\nu = -0.5\pm 0.3$. Like in \eqnref{mbh_forward}, we include the $\pm0.3$ uncertainty by adding random Gaussian numbers $n(\sigma_{\alpha_\nu} = 0.3)$ to mock spectral indices used in the conversion. Then, we magnify these $r$-band magnitudes with randomly drawn magnification values weighted by the smoothed distribution function from \cite{Takahashi_2011}. The simulated population of magnified quasars are those systems with $18.40 < r_{\rm{img}} < 22.12$ and $3 < \mu_{\rm{m}} < 30.5$, which are plotted in \figref{sel_bias}. This selection function corresponds to the image-plane $r$-band magnitudes and magnifications of the brightest WSLQ images.

From this forward-modeling analysis, we obtain an expected bias of $+\selwslq$ in \dmbhmstar for a mock sample of highly magnified quasars at $z \sim 2.1$. For WSLQs and other samples of highly magnified quasars, then, an observed average \dmbhmstar of $+\selwslq$ could be interpreted as consistent with a constant \massrelation relation from $z=2.1$ to $z=0$. The model population of magnified quasars is shown in \figref{sel_bias}, with observed WSLQs plotted over it. There may be a small difference in the \massrelation parameter space between the model and the observed WSLQs, although we note that they are still consistent; each observed WSLQ is located well within $2\sigma$ of the model in \dmbhmstar.

We again qualify that our model for selection bias contains uncertainty yet unaccounted for. We hypothesize that the most prominent sources of uncertainty in these computations are (1) the selection functions themselves, which may be incomplete as presented and utilized here; (2) the possible X-ray\textendash radio dichotomy with regards to AGN selection \citep[e.g.,][]{Aird_2012, Best_2012}; (3) the intrinsic scatter in \Mbh from single-epoch spectroscopy; and for the WSLQ sample (4) the propagation of $r$-band magnitudes given $L_{\rm{bol}}$, as the quasar $K$-corrections contain some uncertainty. (The formula we apply here was originally used for $i$-band magnitudes.) Although the duty cycle is dependent on \Mbh \citep{Schulze_2015}, including this dependence does not significantly impact the measured bias; uncertainties in the GSMF and ERDF that we employ also do not.

\restartappendixnumbering
\section{Gallery of Quasar Spectra and Galaxy SEDs}

We present all quasar spectra used for \Mbh measurements in \figref{gallery_qsospectra} and all host galaxy SEDs modeled with \prospector in \figref{gallery_seds}.

\begin{figure*}
{
\centering
\begin{tabular}{c}
	\includegraphics[width=17cm]{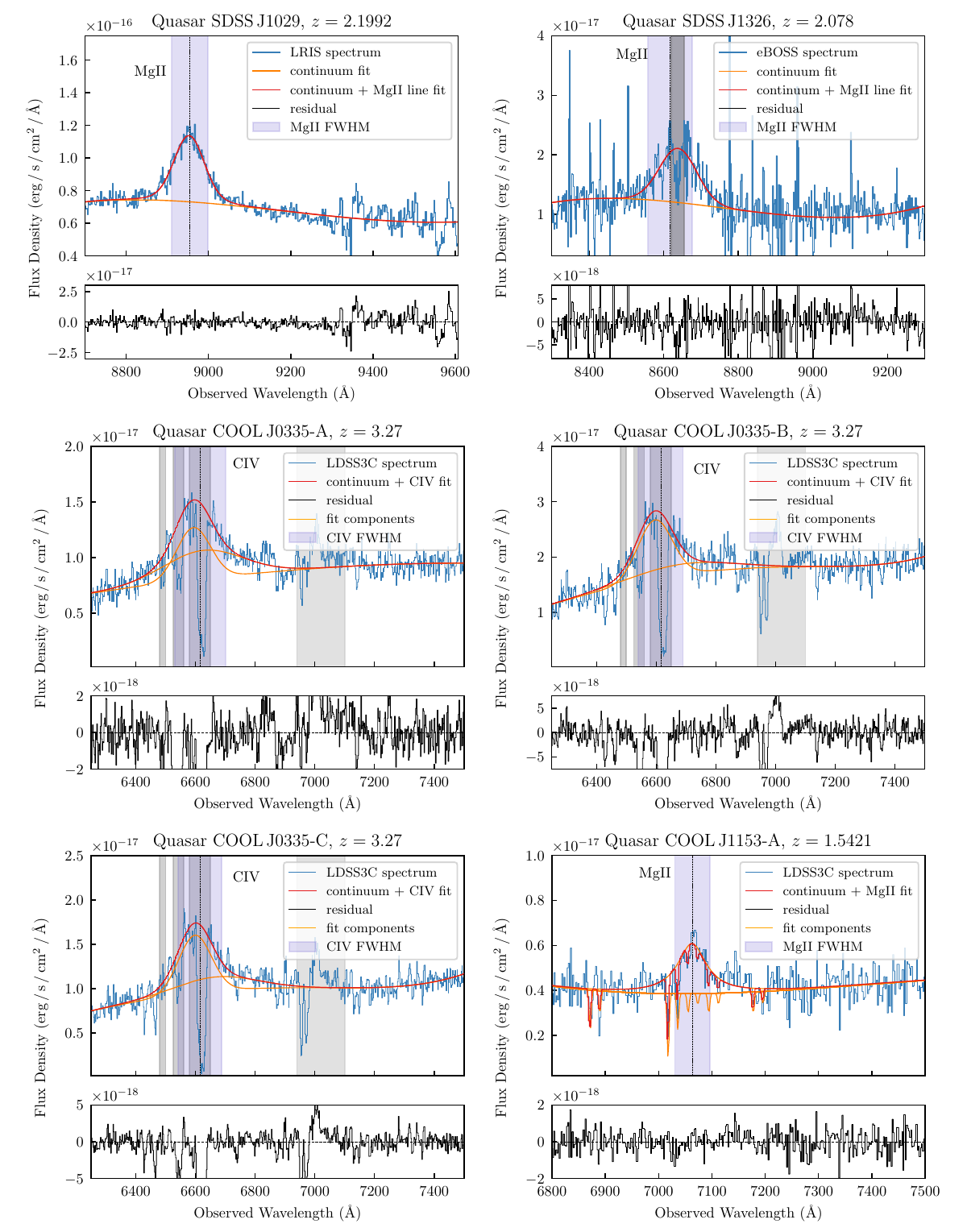}
\end{tabular}
\caption{A gallery of quasar spectra and our best-fit models to observed broad emission lines, from which we derive SMBH masses. Note that we model emission for three distinct quasar images of \sevens, and for \two images of \ones, for the purposes of testing systematics in \secref{imgs_sys}. Different images are labeled as A, B, and C, and they are defined visually in \figref{wslq_images}. For all quasars except for \sevens (where we use \civ), we fit the broad \mgii\ emission line. Regions highlighted in grey are those which are masked when fitting. We mask some spectral features for \fives and \sevens because the underlying emission line can be easily separated, but we fit intervening candidate absorption simultaneously for \eights given the equal spacing and flux-blending with the \mgii\ emission.}
    \label{fig:gallery_qsospectra}
}
\end{figure*}
\begin{figure*}
{
\centering
\begin{tabular}{c}
	\includegraphics[width=17cm]{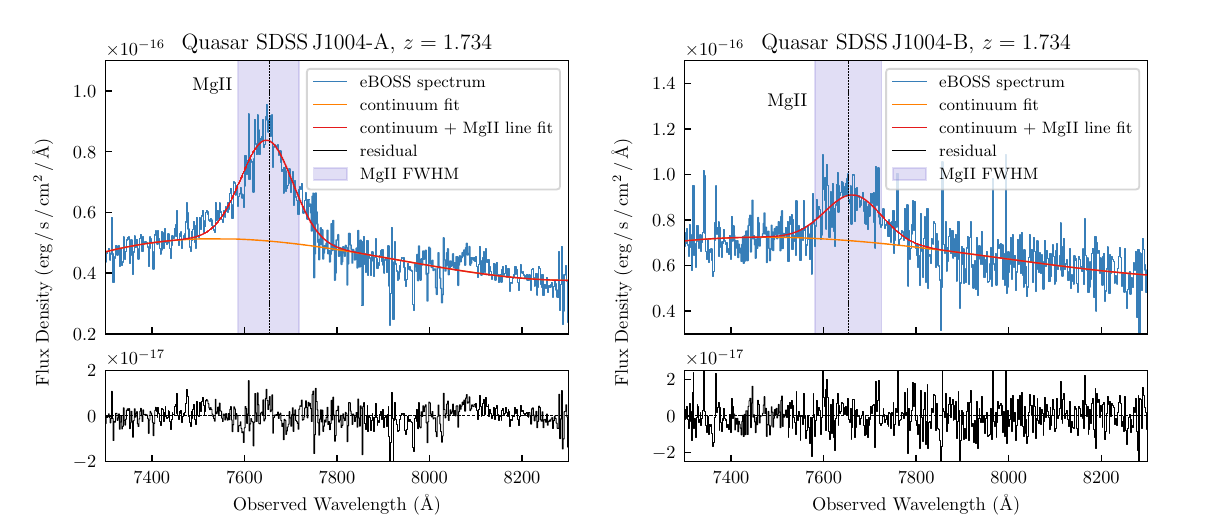}
\end{tabular}
\addtocounter{figure}{-1}
\caption{(continued) A gallery of quasar spectra and our best-fit models to observed broad emission lines, from which we derive SMBH masses. Note that we model emission for three distinct quasar images of \sevens, and for \two images of \ones, for the purposes of testing systematics in \secref{imgs_sys}. Different images are labeled as A, B, and C, and they are defined visually in \figref{wslq_images}. For all quasars except for \sevens (where we use \civ), we fit the broad \mgii\ emission line. Regions highlighted in grey are those which are masked during the fitting. We mask some spectral features for \fives and \sevens because the underlying emission line can be easily separated, but we fit intervening candidate absorption simultaneously for \eights given the equal spacing and flux-blending with the \mgii\ emission.}
    \label{fig:gallery_qsospectra}
}
\end{figure*}

\begin{figure*}
{
\centering
\begin{tabular}{c c}
	\includegraphics[width=8.05cm]{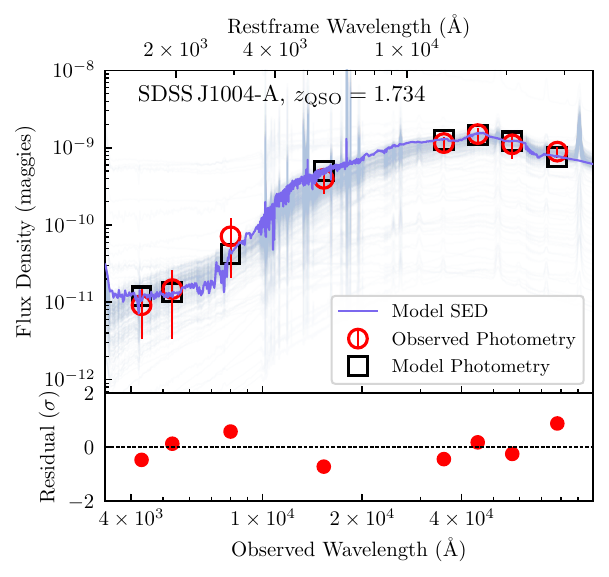} &
	\includegraphics[width=8cm]{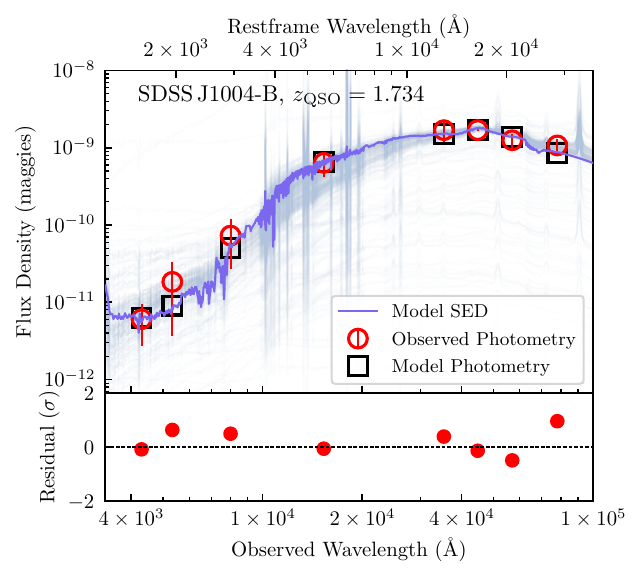} \\
	\includegraphics[width=8cm]{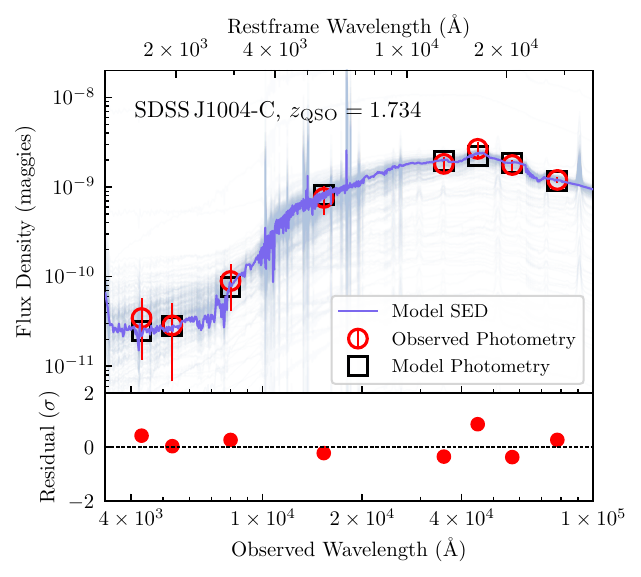} &
	\includegraphics[width=8.3cm]{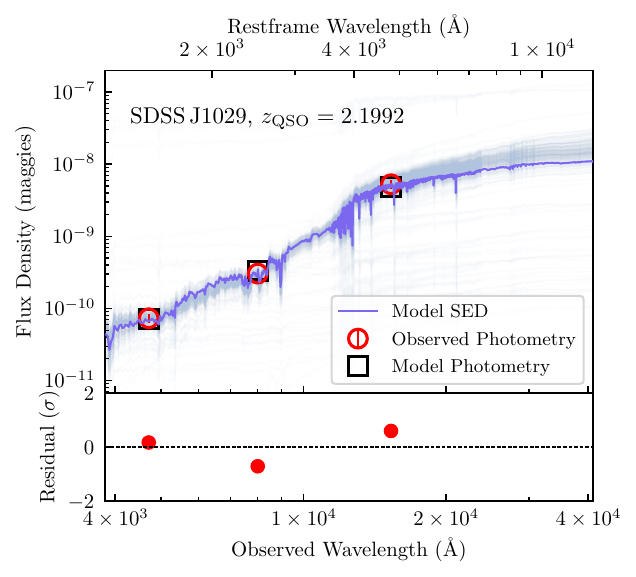} \\
	\includegraphics[width=8cm]{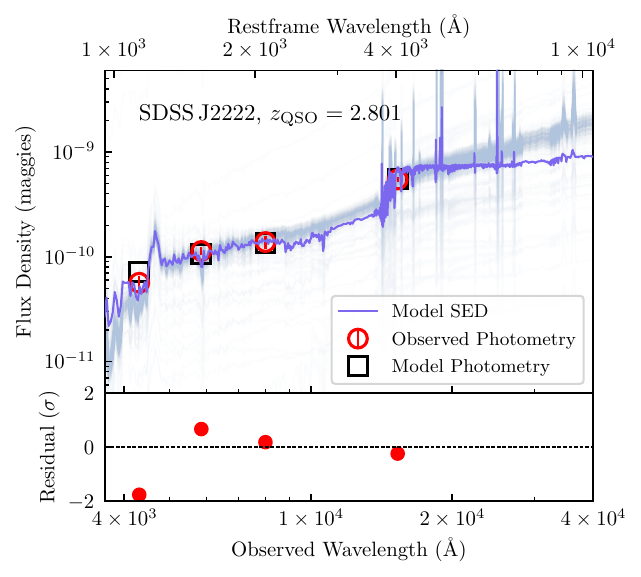} &
	\includegraphics[width=8cm]{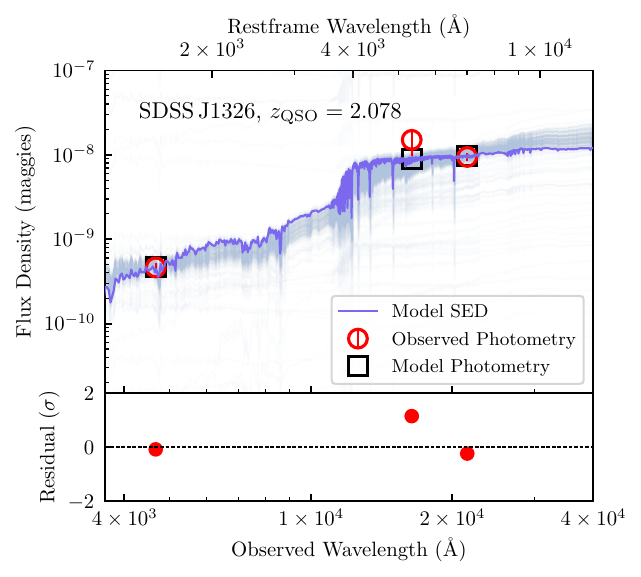}
\end{tabular}
\caption{A gallery of host galaxy SEDs, computed using our SPS modeling framework described in \secref{spsmodeling}. Similar to \figref{gallery_qsospectra}, note that we model three separate images of the quasar host for each of \ones and \eights for testing systematics, with images A, B, and C visually defined in \figref{wslq_images}.}
    \label{fig:gallery_seds}
}
\end{figure*}
\begin{figure*}
{
\centering
\begin{tabular}{c c}
	\includegraphics[width=8cm]{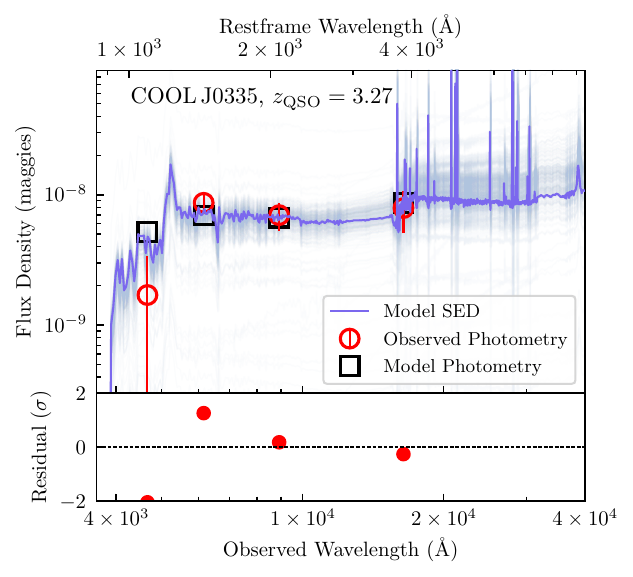} &
	\includegraphics[width=8.1cm]{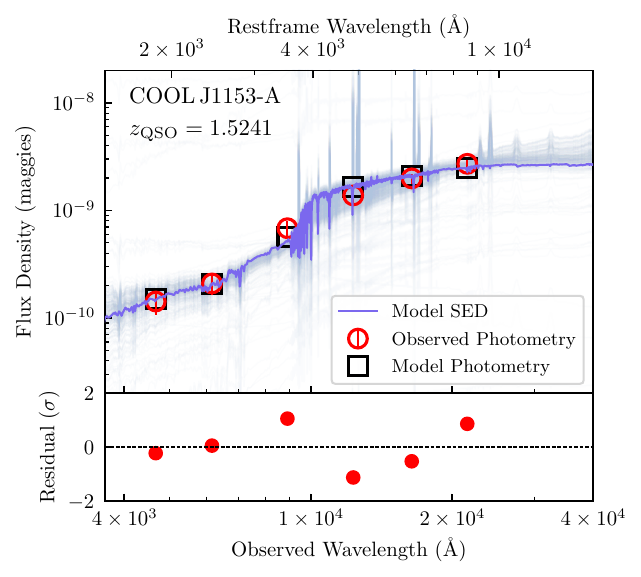} \\
	\includegraphics[width=8.1cm]{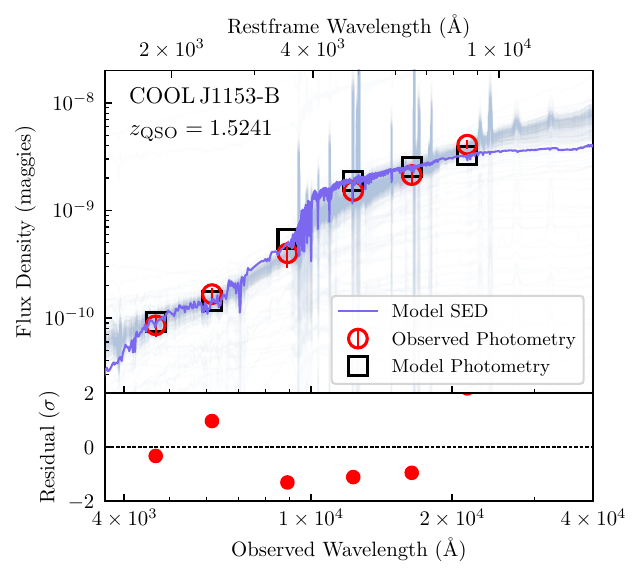} &
	\includegraphics[width=8.1cm]{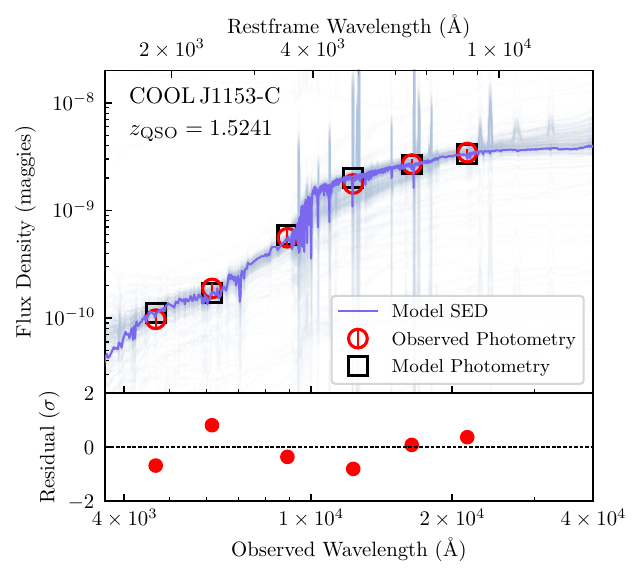}
\end{tabular}
\addtocounter{figure}{-1}
\caption{(continued) A gallery of host galaxy SEDs, computed using our SPS modeling framework described in \secref{spsmodeling}. Similar to \figref{gallery_qsospectra}, note that we model three separate images of the quasar host for each of \ones and \eights for testing systematics, with images A, B, and C visually defined in \figref{wslq_images}.}
    \label{fig:gallery_seds}
}
\end{figure*}

\end{document}